\documentclass[preprint,floatfix]{revtex4}

\pdfoutput=1

\usepackage{pdfpages}
\usepackage{graphicx} 
\usepackage{dcolumn} 
\usepackage{amsmath}
\usepackage{amssymb}
\usepackage{bm}
\usepackage{times}
\usepackage{ulem}
\usepackage[justification=raggedright]{caption}

\usepackage{subfigure}
\usepackage{tabularx}
\usepackage{appendix}

\usepackage{amstext}
\usepackage{array}

\usepackage{bbm}

\definecolor{wine-stain}{rgb}{0.5,0,0} 
\definecolor{bblue}{rgb}{0,0.0,0.5} 

\usepackage[colorlinks=true
,urlcolor=blue
,anchorcolor=blue
,citecolor=blue 
,filecolor=blue
,linkcolor=wine-stain
,menucolor=blue
,pagecolor=blue
,linktocpage=true
,pdfproducer=medialab
,linktoc=all 
]{hyperref}

\usepackage{booktabs}  

\allowdisplaybreaks

\newcommand{\ncmd}{\newcommand}

\ncmd{\lt}{\left}
\ncmd{\rt}{\right}
\ncmd{\tr}[1]{~\mbox{tr}\lt\{ {#1}\rt\}}
\ncmd{\half}{\frac{1}{2}}
\ncmd{\eps}{\epsilon}
\ncmd{\veps}{\varepsilon}
\ncmd{\dgr}{\dagger}
\ncmd{\sig}{\sigma}
\ncmd{\gam}{\gamma}
\ncmd{\rtarw}{\rightarrow}
\ncmd{\Rt}{\Rightarrow}
\ncmd{\abs}[1]{\lt\cb{#1}\rt\cb}
\ncmd{\avg}[1]{\lt\lb{#1}\rt\rb}
\ncmd{\dl}{\delta}
\ncmd{\Dl}{\delta}
\ncmd{\sgn}[1]{\mbox{sgn}\lt(#1\rt)}
\ncmd{\kap}{\kappa}
\ncmd{\wtil}[1]{\widetilde{#1}}
\ncmd{\thrfr}{\therefore}
\ncmd{\eq}[1]{Eq. \eqref{#1}}
\ncmd{\fig}[1]{Fig. \ref{#1}}
\ncmd{\Lam}{\Lambda}
\ncmd{\lam}{\lambda}
\ncmd{\dow}{\partial}
\ncmd{\ordr}[1]{\mathcal{O}\lt(#1\rt)}
\ncmd{\dsty}{\displaystyle}
\ncmd{\alert}[1]{\color{red}{#1}}
\ncmd{\mc}{\mathcal}
\ncmd{\mbf}[1]{\mathbf{#1}}
\ncmd{\Deriv}[2]{\frac{d{#1}}{d{#2}}}
\ncmd{\ParDeriv}[2]{\frac{\partial{#1}}{\partial{#2}}}
\ncmd{\step}[1]{\Theta\lt(#1\rt)}
\ncmd{\td}{\tilde} 
\ncmd{\what}{\widehat}
\ncmd{\om}{\omega}
\ncmd{\Om}{\Omega}
\ncmd{\vrho}{\varrho}
\ncmd{\vsig}{\varsigma}
\ncmd{\vkap}{\varkappa}
\ncmd{\bqa}{\begin{eqnarray}} 
\ncmd{\eqa}{\end{eqnarray}}
\ncmd{\nn}{\nonumber \\}
\ncmd{\nnum}{\nonumber}
\ncmd{\comment}[1]{{\color{red}{#1}}}
\definecolor{new_color}{RGB}{50,155,0}

\ncmd{\lb}{\big<}
\ncmd{\rb}{\big>}
\ncmd{\cb}{\big|}
\ncmd{\cH}{{\cal H}}
\ncmd{\lc}{l_c}
\ncmd{\Lc}{l_c}

\ncmd{\cphir}{\cb \phi \rb}
\ncmd{\lphic}{\lb \phi \cb}

\ncmd{\ctr}{\cb T \rb}
\ncmd{\ltc}{\lb T \cb}
\ncmd{\cpr}{\cb p \rb}
\ncmd{\lpc}{\lb p \cb}

\ncmd{\cTr}{\cb {\cal T} \rb}

\ncmd{\ccr}{\cb \chi \rb}
\ncmd{\lcc}{\lb \chi \cb}

\ncmd{\cS}{{\cal S}}
\ncmd{\tcS}{\tilde { \cal S}}

\ncmd{\sqg}{\sqrt{|g(x)|}}
\ncmd{\gmn}{E_{\mu i}}

\ncmd{\emi}{E_{\mu i}}
\ncmd{\emj}{E_{\mu j}}
\ncmd{\eni}{E_{\nu i}}
\ncmd{\enj}{E_{\nu j}}
\ncmd{\de}{|E(x)|}

\ncmd{\gemn}{g_{E,\mu \nu}}
\ncmd{\gemnp}{g_{E',\mu \nu}}

\ncmd{\bgmn}{\bar E_{\mu i}}
\ncmd{\Fs}{F(\sigma)}
\ncmd{\cJ}{ {\cal J}\left(E',\sigma';E,\sigma \right)  }
\ncmd{\Si}{ \Sigma \left(E',\sigma';E,\sigma \right)  }

\ncmd{\grs}{g_{\alpha \beta}}
\ncmd{\pmn}{\pi^{\mu i}}
\ncmd{\prs}{\pi^{\alpha \beta}}
\ncmd{\fs}{ \frac{3}{\sqrt{2}} \sigma }
\ncmd{\fst}{ 3 \sqrt{2} \sigma }

\ncmd{\ee}{entanglement entropy }

\ncmd{\pppi}{\left( \phi_{i_1}   \phi_{i_2}  ...  \phi_{i_{2n}} \right)}
\ncmd{\ppp}{O_{ i_1 i_2 .. i_{2n}} }
\ncmd{\TTT}{T_{i_1 i_2 .. i_{2n}}}
\ncmd{\PPP}{P_{i_1 i_2 .. i_{2n}}}
\ncmd{\ttt}{t_{i_1 i_2 .. i_{2n}}}
\ncmd{\eee}{e_{i_1 i_2 .. i_{2n}}}

\ncmd{\cT}{{\cal T} }
\ncmd{\cP}{{\cal P} }
\ncmd{\cV}{{\cal V} }
\ncmd{\cW}{{\cal W} }

\ncmd{\ccw}{\cos( 2 \omega \delta) }
\ncmd{\ssw}{\sin (2 \omega \delta) }

\makeatletter
\newcommand*{\rom}[1]{\expandafter\@slowromancap\romannumeral #1@}
\makeatother

\begin{document}

\title{
State dependent spread of entanglement in relatively local Hamiltonians 
}

\author{Sung-Sik Lee\\
\vspace{0.3cm}
{\normalsize{Department of Physics $\&$ Astronomy, McMaster University,}}\\
{\normalsize{1280 Main St. W., Hamilton ON L8S 4M1, Canada}}
\vspace{0.2cm}\\
{\normalsize{Perimeter Institute for Theoretical Physics,}}\\
{\normalsize{31 Caroline St. N., Waterloo ON N2L 2Y5, 
Canada}}
}

\date{\today}

\begin{abstract}

Relatively local Hamiltonians are a class of background independent non-local Hamiltonians
from which local theories emerge within a set of short-range entangled states.
The dimension, topology and geometry of the emergent local theory 
is determined by the initial state to which the Hamiltonian is applied.
In this paper, we study dynamical properties 
of a simple relatively local Hamiltonian
for $N$ scalar fields in the large $N$ limit.
It is shown that the coordinate speeds at which entanglement spreads
and local disturbance propagates in space strongly
depend on state in the relatively local Hamiltonian.

\end{abstract}

\maketitle


\newpage

{
\hypersetup{linkcolor=bblue}
\hypersetup{
    colorlinks,
    citecolor=black,
    filecolor=black,
    linkcolor=black,
    urlcolor=black
}
\tableofcontents
}


\section{Introduction}

Locality, one of the cherished principles in quantum field theory\cite{PhysRev.132.2788},
is unlikely to be a part of
the yet unknown fundamental theory of nature that includes dynamical gravity\cite{PhysRevLett.114.031104}.
In the presence of strong quantum fluctuations of metric,
there is no well-defined notion of what is near and what is far.
At the same time, a quantum theory of gravity 
that reproduces the general relativity in the classical limit 
can not be a generic non-local theory either.
If quantum fluctuation of geometry is weak,
a local field theory should emerge 
as an effective description of quanta
propagating on top of the classical geometry
that describes a saddle point 
of the fundamental theory.
In this case, locality defined with respect to the saddle point geometry
is determined by the state not by the fundamental Hamiltonian.
One possibility is that the microscopic theory that includes gravity belongs 
to a class of non-local theories 
which  act approximately as local theories
within a set of states that represent classical geometry.
Such theories, while being non-local in the usual sense, must have 
a weaker sense of locality. 
We call it {\it relative locality}
that locality of Hamiltonian 
is determined relative to state.

Recently, it has been argued that
the general relativity can arise
as a semi-classical description 
of a relatively local quantum theory
of matter fields\cite{Lee2018}.
In the construction, spatial metric is introduced as a collective variable
that characterizes how matter fields are entangled in space 
within a sub-Hilbert space.
One can design a Hamiltonian for the matter fields 
such that it induces the classical general relativity 
for the collective variable  at long distances
in the limit that the number of matter fields is large. 
The Hamiltonian is shown to have the property that
the range of interactions 
depends on the entanglement present in states 
on which the Hamiltonian acts.
Although the theory is non-local in the strict sense,
it differs from generic non-local theories\cite{PhysRevLett.70.3339,Kitaev_SYK}
in that a local effective theory emerges in a space of short-range entangled states. 
In the theory, the notion of distance that sets locality 
of Hamiltonian is determined by entanglement present in state.
In relatively local theories,
the relation between entanglement and geometry\cite{
PhysRevLett.96.181602,
1126-6708-2007-07-062,
VanRaamsdonk:2010pw,
Casini2011,
Lewkowycz2013}
uncovered in the context of AdS/CFT correspondence\cite{
Maldacena:1997re,Witten:1998qj,Gubser:1998bc} 
is promoted to a kinematic principle 
that posits that
 geometry is a collective degree of freedom 
that encodes  entanglement of matter fields. 
In this sense, 
relatively local theories can be viewed as 
examples of emergent gravity
that realize the $ER=EPR$ conjecture\cite{doi:10.1002/prop.201300020}.

However, the relatively local theory introduced in Ref. \cite{Lee2018} is defined only perturbatively
in the semi-classical limit.
The construction starts with matter fields defined on a manifold with a fixed dimension and topology,
and it is not capable of describing non-perturbative processes.
Furthermore, the continuum Hamiltonian of the matter fields is rather complicated,
and its dynamics hasn't been studied explicitly.
Therefore it is of interest 
to consider tractable relatively local models
that can be defined beyond the perturbative level.
In this paper, we construct a simple relatively local Hamiltonian for $N$ scalar fields, 
and study its dynamics in the large $N$ limit.

Here is the outline of the paper.
In Sec. II, we define the Hilbert space and the Hamiltonian.
Sec. II A starts with the Hilbert space of $N$ scalar fields 
which are defined on a collection of sites.
Within the full Hilbert space, we focus 
on a sub-Hilbert space that is invariant 
under an internal symmetry group. 
The symmetric sub-Hilbert space is spanned by basis states
which are labeled by multi-local singlet collective variables.
In Sec. II B we introduce a relatively local Hamiltonian
that is invariant not only under the internal symmetry
but also under the full permutation group of the sites.
The relatively local Hamiltonian has no preferred background,
and is non-local as a quantum operator.
However, it acts as a local Hamiltonian  
within a set of states with local structures of entanglement in the large $N$ limit.
A state is said to have a local structure of entanglement 
(henceforth called `local structure', in short)
if entanglement entropy of any sub-region 
can be written as the volume of its boundary
measured with a classical notion of distance (metric)
associated with a lattice (manifold).
The existence of such a lattice, and the dimension, topology and geometry of the lattice, if exists,
are all determined by the pattern of entanglement of  state.
The lattice associated with a state with a local structure 
consequently determine the dimension, topology and geometry 
of the local theory that emerges when the Hamiltonian is applied to that state.
In this sense, the locality of the Hamiltonian is set by states.
In Sec. III, we study time evolution of states with local structures.
We derive an induced Hamiltonian for the collective variables
that governs the time evolution of states 
within the symmetric sub-Hilbert space.
In the large $N$ limit, the time evolution is described by
classical equations of motion for the collective variables.
Sec. III A is devoted to the perturbative analysis 
that describes time evolution of states 
which are close to direct product states.
It treats small entanglement as a perturbation added
to an unentangled ultra-local state.
It is shown that the coordinate speed at which entanglement spreads
in space vanishes in the limit that the entanglement of the initial state vanishes. 
In Sec. III B, we  numerically integrate the classical equation of motion 
for the collective variables.
We confirm that the coordinate speed at which entanglement spreads
depends strongly on the amount of entanglement in the initial state.
Furthermore, it is shown that the local structure of the initial state
determines the dimension of the emergent local theory.
In Sec. IV, we examine propagation of 
a local disturbance added to a translationally invariant initial state.
It is shown that a small local disturbance propagates on top of the geometry 
set by the collective variables that are formed in the absence of the disturbance. 
The coordinate speed of the propagating mode is determined by the initial state 
because the geometry is set by the state.

\section{Model}

\subsection{Hilbert space}

 \begin{figure}[ht]
 \begin{center}
 \includegraphics[scale=0.5]{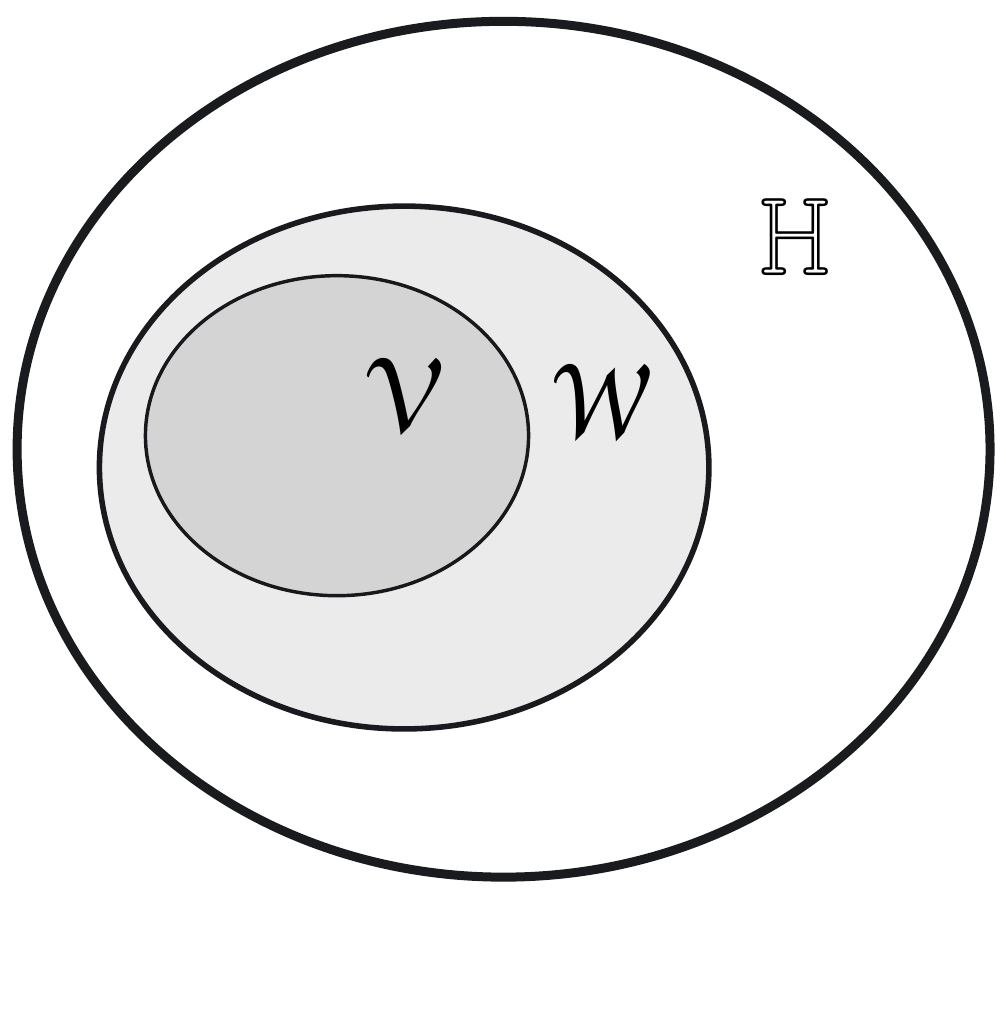} 
 \end{center}
 \caption{
 ${\mathbb H}$ is the full Hilbert space of the scalar fields.
 $\cW$ is the $S_N \ltimes Z_2^N$ invariant sub-Hilbert space of   ${\mathbb H}$.
  $\cV$ is the $O(N)$ invariant sub-Hilbert space of   $\cW$.
  }
 \label{fig:Hilbertspace}
 \end{figure}

We first introduce the Hilbert space in which our model is defined.
We consider a quantum system defined on a set of sites labeled by $i=1,2,..,L$.
At each site, there are $N$ real scalar fields $\phi_i^a$ with $a=1,2,..,N$.
The full Hilbert space is spanned by the basis states 
$\Bigl\{ \cb \phi \rb \cb  -\infty < \phi^a_i < \infty \Bigr\}$,
where $ \cb \phi \rb $ is the eigenstate of the field operator
that satisfies $\hat \phi^a_i \cb \phi \rb = \phi^a_i \cb \phi \rb$
and $\lb \phi' \cb \phi \rb = \prod_{i,a} \delta \left(
\phi_i^{'a} - \phi_i^a
\right)$.
Within the full Hilbert space, 
we consider the
 $S_N \ltimes Z_2^N$ invariant
 sub-Hilbert space denoted as ${\cal W}$.
Here $S_N$ is the global flavor permutation group
under which the field transforms as 
$\phi_i^a \rightarrow \phi_i^{P_a}$.
Under each of the $N$ independent $Z_2$ transformations, 
one component of the scalar field flips sign as
 $\phi_i^a \rightarrow (-1)^{\delta_{ab}} \phi_i^a$
 with $b=1,2,..,N$.
$S_N \ltimes Z_2^N$  is a subgroup of the $O(N)$ group.
General wavefunctions in $\cW$ can be written as functions of 
multi-local single-trace operators with even powers, 
\bqa
\ppp \equiv \frac{1}{N} \sum_a 
\phi^a_{i_1} \phi^a_{i_2} ... \phi^a_{i_{2n}}.
\label{eq:multi}
\eqa
This is because $\{ \ppp \}$ forms a complete basis 
for the polynomials of the fundamental fields
that are invariant under $S_N \ltimes Z_2^N$.
Consequently,  ${\cal W}$ can be spanned by basis states,  
\bqa
\cb  {\cal T} \rb = 
\int D \phi ~ 
e^{-  N \sum_{n} \TTT  \ppp  }   \cphir,
\label{t}
\eqa
where $D \phi \equiv \prod_{i,a} d \phi^a_i$.
Henceforth, repeated site indices are understood to be summed over all
 sites unless mentioned otherwise.
The set of symmetric complex  tensors of even ranks,
${\cal T} = \Bigl\{ T_{ij}, T_{ijkl}, ... \Bigr\}$ 
are collective variables that 
label basis states of $\cW$.
The subset of states with pure imaginary collective variables,
$ \Bigl\{ \cb {\cal T} \rb ~ | ~ \TTT = i \ttt,
 ~ \ttt \in \mathbb{R} \Bigr\}$
is already a complete basis of ${\cal W}$.
This follows from the fact that $\ttt$ is the Fourier conjugate variable of $\ppp$.
Therefore, general states in ${\cal W}$ can be expressed as
\bqa
\ccr  = \int D {\cal T} ~ \cb {\cal T} \rb \chi({\cal T}),
\label{ccr}
\eqa
where 
$D {\cal T} \equiv \prod_{i \geq j} d T_{ij} \prod_{i \geq j \geq k \geq l} d T_{ijkl} ... $,
and the integration over $\TTT$ is defined along the imaginary axis.
$\chi({\cal T})$ is a wavefunction for the collective variables.
Although it is defined on the imaginary axes of ${\TTT}$'s,
it is useful to analytically extend  $\cb {\cal T} \rb$ and $\chi({\cal T})$ 
to the complex planes for the following reasons.
First, $\cTr$ is not normalizable if $\TTT$ is pure imaginary,
and one should add a real symmetric tensor $\eee$ to
construct normalizable states. 
Real components of the tensors create normalizable wave-packets. 
Second, $\chi({\cal T})$  has saddle points off the imaginary axes in general,
and one needs to deform the path of $\int d \TTT$ in the complex planes
for the saddle-point approximation.
Within ${\cW}$
there is a sub-space $\cV$
that is $O(N)$ symmetric.
Among all multi-local single-trace operators in \eq{eq:multi}, 
only bi-local operators are invariant under the $O(N)$ transformations.
Therefore, $\cV$ is spanned by $\cb \cT \rb$
with $\TTT=0$ for $n\geq2$.
We denote basis states of $\cV$ as $\ctr$
which is labeled by $T_{ij}$ only.
This is illustrated in \fig{fig:Hilbertspace}.
$\cV$ will be the main focus of the present work. 

 \begin{figure}[ht]
 \begin{center}
 \includegraphics[scale=0.5]{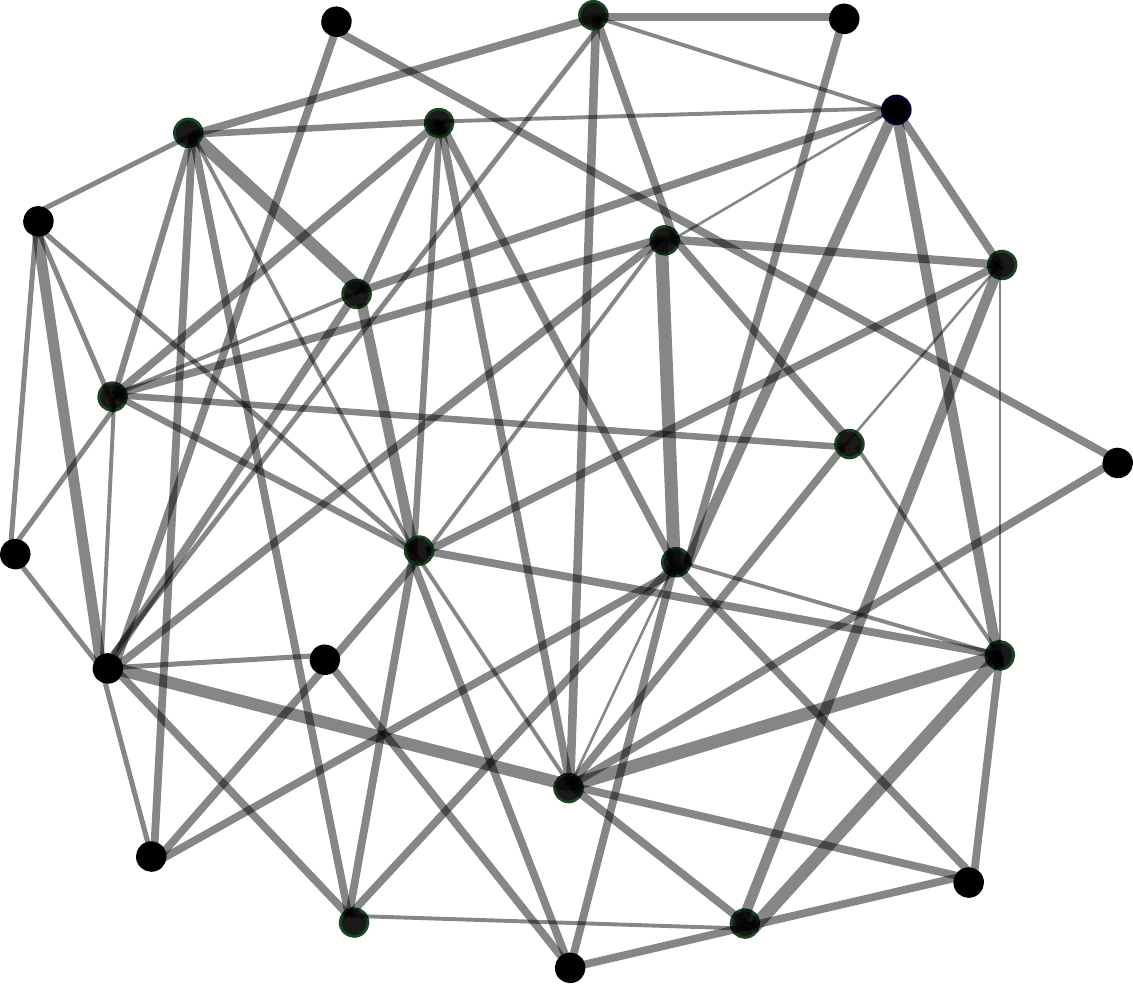} 
 \end{center}
 \caption{
A set of sites connected by bi-local collective variables,
where the thickness of the bond between $i,j$ represents the magnitude of $T_{ij}$.
For generic choices of bi-local fields, graphs do not represent regular lattices. 
}
 \label{fig:graph}
 \end{figure}

%
The collective variables are a direct measure of entanglement.
For a state  $\ctr$ in $\cV$
with  $|T_{i \neq j}| \ll e_{ii}, e_{jj} $  
where $T_{ij} = e_{ij} + i t_{ij}$,
the entanglement entropy between a subset of sites $A$ and its compliment $\bar A$ 
is given by\cite{Lee2016}
\bqa
S_A = N 
\left[
\sum_{i \in A, j \in \bar A} 
 \left(
- \ln \frac{| T_{ij}|^2}{4 e_{ii} e_{jj}}
+ 1
\right)
\frac{ |T_{ij}|^2}{4 e_{ii} e_{jj}}
+  O\left( (T/e)^4 \right)
\right]
+ O(N^0).
\eqa
To the second order in $T_{ij}$,
the mutual information between sites $i$ and $j$
is 
$I_{ij} 
=
N 
 \left(
- \ln \frac{| T_{ij}|^2}{4 e_{ii} e_{jj}}
+ 1
\right)
\frac{ |T_{ij}|^2}{4 e_{ii} e_{jj}}
+ O(N^0)
$ 
\footnote{
The entanglement entropy 
of related Gaussian wavefunctions has been 
also obtained in Ref. \cite{Mozaffar2016}.
}.
It is useful to visualize the mutual informations between sites
in terms of entanglement bonds as is shown in \fig{fig:graph}.
A notion of distance between sites 
can be defined based on entanglement\cite{PhysRevD.95.024031}.
For example, one can define a distance between sites $i$ and $j$
as a function of the bi-local fields such that
sites that can be connected by few strong bonds
have a small proper distance,
and sites that can only be connected by
many weak bonds have a large proper distance.
One such distance function
that captures this idea in the limit that $|T_{ij}|$ is small is 
 $\min_C \left[ -\sum_{(l,m) \in C} \ln |T_{lm} |^2 \right]$
where $C$ denotes chains of bi-local fields 
that connect $i$ and $j$.
%
%

 \begin{figure}[ht]
 \begin{center}
   \subfigure[]{
 \includegraphics[scale=0.5]{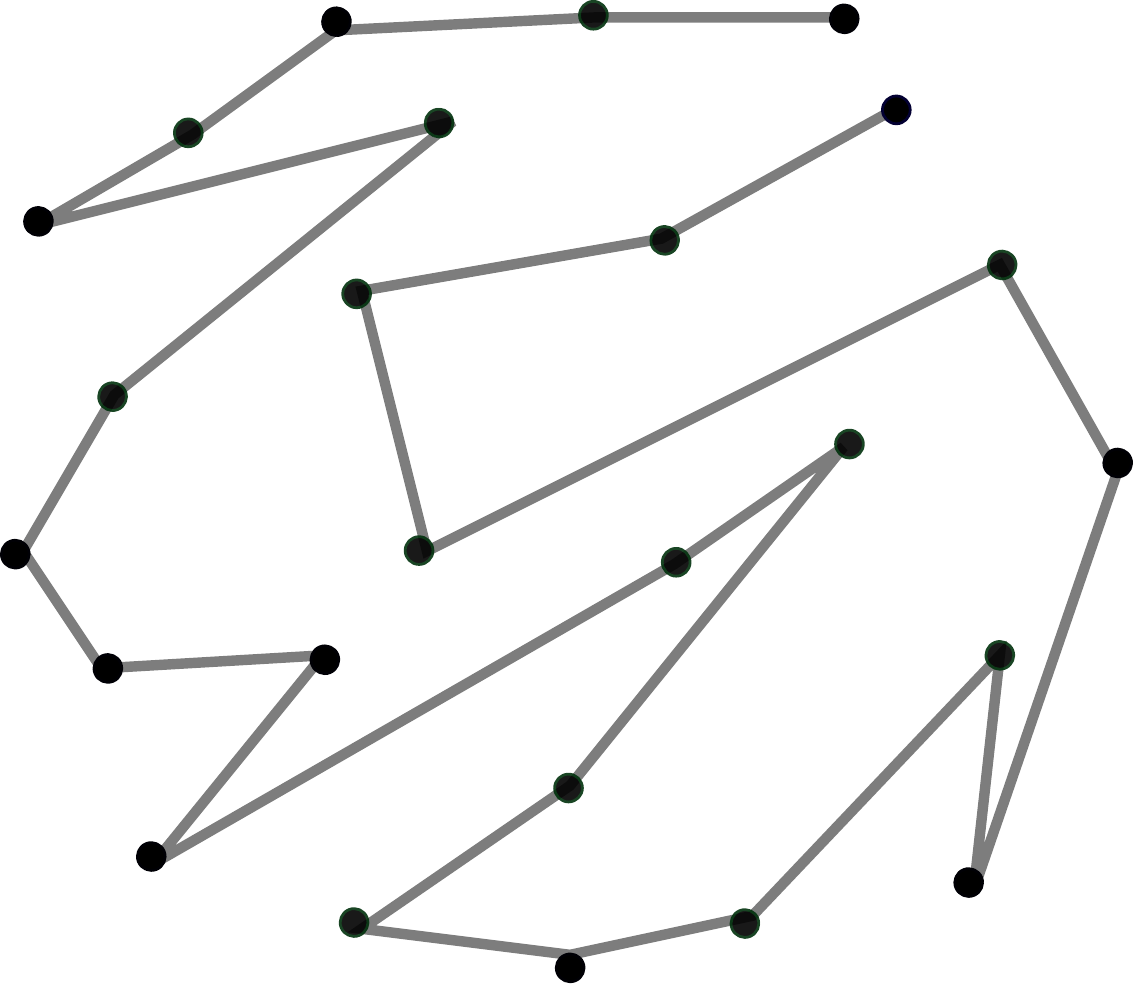} 
 } 
 \hfill
 \subfigure[]{
 \includegraphics[scale=0.5]{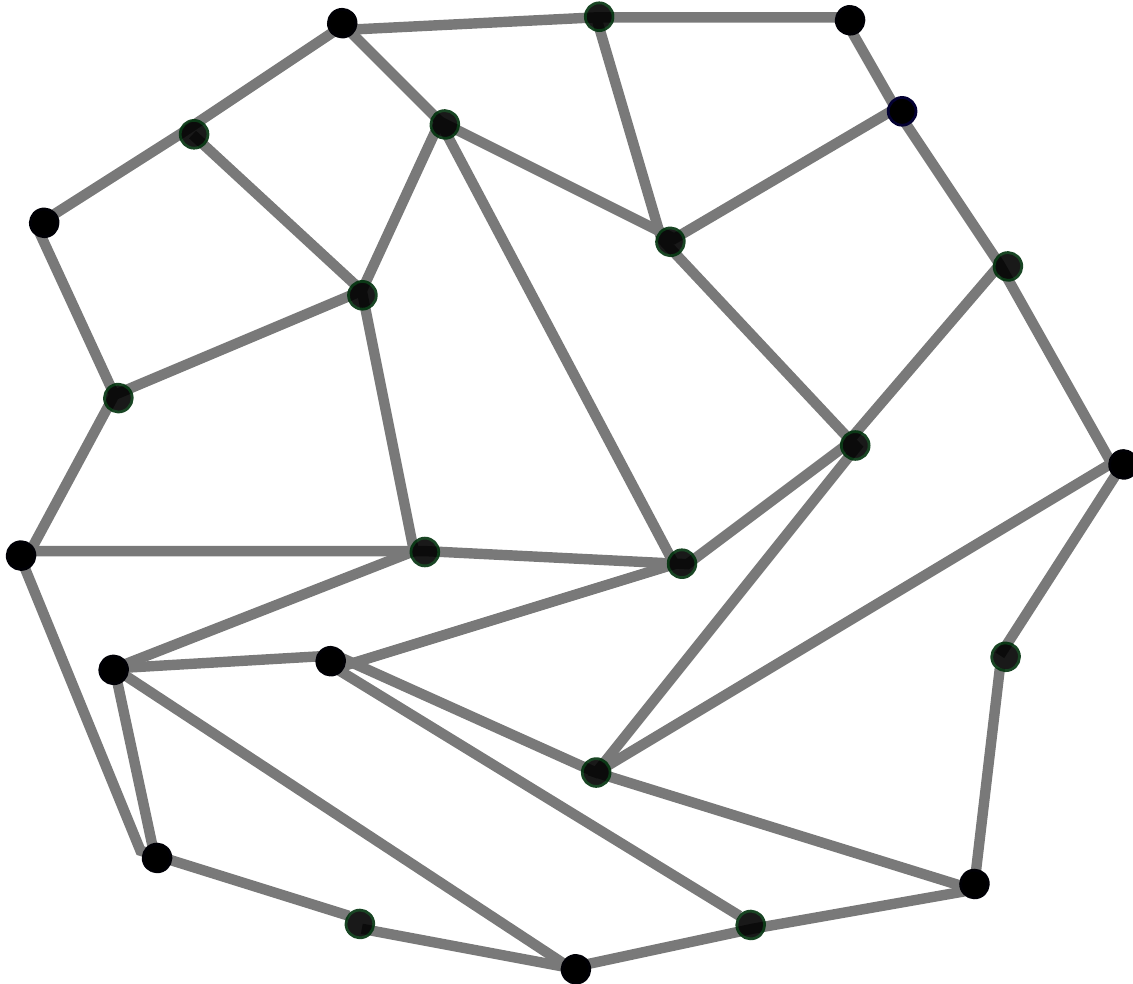} 
 } 
  \end{center}
 \caption{
 Entanglement bonds that exhibit
 one and two dimensional local structures, respectively.
 }
 \label{fig:graph_12D}
 \end{figure}

Graphs that emerge from 
random choices of $T_{ij}$
do not represent any lattice in general.
There exists a special set of states 
whose entanglement bonds exhibit 
a lattice with a sense of locality.
We define that a state has a {\it local structure}
if the entanglement entropy of any sub-region
is proportional to the volume of its boundary 
defined with respect to a lattice.
Whether there exists such lattice or not is a property of states.
A local structure arises 
if the pattern of entanglement 
exhibits a sense of locality.
For example, 
$T_{ij}(0) =  \delta_{ij} + \epsilon \delta_{ d_{ij}^{(1)}, 1}$
with $d_{ij}^{(1)} = [i - j]_L$ 
and $[x]_L  = min \left\{ |x + n L|  \Big| n \in \mathbb{Z} \right\}$
describes the one-dimensional lattice with nearest neighbor bonds
and the periodic boundary condition.
A torus of $\sqrt{L} \times \sqrt{L}$ square lattice arises in states with
$T_{ij}(0) = \delta_{ij} + \epsilon  \delta_{ d_{ij}^{(2)}, 1}$
and
$d_{ij}^{(2)} = 
\sqrt{
\left[ i - j  \right]_{\sqrt{L}}^2 
+
\left[ 
\left\| \frac{i-1}{\sqrt{L}} \right\|
- 
\left\| \frac{j-1}{\sqrt{L}} \right\|
\right]_{\sqrt{L}}^2
}$,
where $\left\| x \right\|$ represents the largest integer equal to or smaller than $x$.
The entanglement bonds with one and two dimensional local structures
are shown in \fig{fig:graph_12D}.
Here, the dimension, topology and geometry of the emergent lattice are properties of states
determined by the collective variable.
In the thermodynamic limit, 
${\cal V}$ includes states that describe any lattice in any dimension.
%

\subsection{Relatively Local Hamiltonian}

Now we construct a relatively local Hamiltonian
whose dimension, topology and locality are determined by those of states.
For this, we can not have a local hopping term that is based on a pre-determined notion of distance.
Instead, the range and the strength of hopping need to be state dependent.
Here we consider a simple relatively local Hamiltonian that is invariant under $S_N \ltimes Z_2^N$,
\bqa
\hat H = -R \sum_{i,j} \sum_a 
\hat T_{ij}^a  ( \hat \phi_i^a  \hat \phi_j^a ) 
 + U \sum_i \sum_a \hat \pi_i^a  \hat \pi_i^a 
+ \frac{ \lambda }{N}  \sum_i \sum_{a,b}  ( \hat \phi_i^a  \hat \phi_i^a ) ( \hat \phi_i^b  \hat \phi_i^b ).
\label{Hhat}
\eqa
Here 
$R$, $U$, $\lambda$
are constants,
and $\hat \pi_i^a$ is the conjugate momentum of $\hat \phi_i^a$.
The second and the third are the standard 
kinetic and quartic interaction terms 
that are ultra-local.
The first term describes `hoppings'. 
If the hopping was nonzero 
only between nearby sites in a fixed lattice, 
this Hamiltonian would be a usual local Hamiltonian.
The key differences of \eq{Hhat} 
from such local Hamiltonians  are that 
$i,j$ run over all pairs of sites,
and the hopping amplitude is an operator.
The strength of hopping between two sites
is determined by $\hat T_{ij}^a$ which 
measures inter-site entanglement bonds.
Since there can be potentially nonzero coupling between any two sites,
$\hat H$ is a non-local Hamiltonian as a quantum operator.
However,  $\hat T_{ij}^a$ can be chosen such that $\hat H$ acts as a local Hamiltonian
within states that have local structures.

Because $T_{ij}$ is the collective variable that determines pairwise connectivity, 
one may try to construct $\hat T_{ij}^a$ such that
$
\hat T_{ij}^a \ctr = T_{ij} \ctr$.
However, it is not easy to find such operator 
because not all states with different collective variables 
are linearly independent for $L \gg N$\footnote{
This follows from the fact that there are more collective variables
than the fundamental fields in the thermodynamic limit
with a fixed $N$.}.
Alternatively, we look for an operator 
of which $\ctr$ is an approximate eigenstate.
One such operator that commutes with
$\hat \phi_i^a \hat \phi_j^a$ is
\bqa
\hat T_{ij}^a =\frac{1}{N-1} \sum_{b \neq a}  \hat \pi_i^b  \hat \pi_j^b.
\eqa
For states in $\cV$, 
the operator satisfies
\bqa
\hat T_{ij}^a \ctr = \int D \phi ~
\left[
2 T_{ij} - 4 T_{ik} T_{lj} \frac{ \sum_{b \neq a} \phi_k^b  \phi_l^b}{N-1}
\right]
e^{- N T_{ij} O_{ij} }   \cphir.
\label{eq:TaT}
\eqa
While $\ctr$ is not an eigenstate of $\hat T_{ij}$,
its expectation value is given by
\bqa
\frac{ \lb T \cb \hat T_{ij} \cb T \rb}{\lb T \cb T \rb}
= 2  (T^{*-1} + T^{-1})^{-1}_{ij}.
\label{hatTT}
\eqa
Here $T_{ij}^{-1}$ denotes the inverse of $T_{ij}$ 
as an $L \times L$ matrix.
\eq{hatTT} reduces to $T_{ij}$ for real $T_{ij}$.
$\hat T_{ij}^a$ is an operator that measures
the strength of the bond between $i$ and $j$.
Therefore, $-\hat T_{ij}^a \hat \phi_i^a \hat \phi_j^a$
represents hoppings between all pairs of sites 
with strength proportional to the 
pre-existing entanglement between the sites.
It is noted that \eq{Hhat} has the full permutation symmetry of $L$ sites.
The permutation group includes the usual discrete translational symmetry
of any regular lattice as subgroups, 
and the Hamiltonian does not have a preferred background.

$\hat T_{ij}$ is a quantum operator that measures  the geometry of states.
In quantum mechanics, it is in general impossible to read information 
without disturbing the state\cite{Busch1996}.
However, $\hat T_{ij}$ acts as a classical variable
 to the leading order in $1/N$
when applied to semi-classical states.
This is attributed to the fact that
the geometric information is encoded redundantly 
by a large number of matter fields
in the large $N$ limit.
To see this,  we consider a general state in 
\eq{ccr} with a semi-classical wavefunction for the collective variables,
\bqa
\chi(\cT) & = & e^{ 
\sum_n
\left[ 
N  \bar P_{i_1 i_2..i_{2n}}  \left( \TTT -  \bar T_{i_1 i_2..i_{2n}} \right)  + \frac{ \left( \TTT -  \bar T_{i_1 i_2..i_{2n}} \right)^2 }{2 \Delta^2} 
\right]
},
\label{cct}
\eqa
where $  \bar T_{i_1 i_2..i_{2n}} $ and $  \bar P_{i_1 i_2..i_{2n}}$ denote
classical collective variables and their conjugate momenta, respectively.
\eq{cct} is a normalizable wavefunction with width $\Delta$ defined
along the imaginary axes of $ T_{i_1 i_2..i_{2n}}$'s.
For $1/N \ll \Delta \ll 1$, both the collective variables
and their conjugate momenta are well defined.
For simplicity, let us consider states with 
$\bar T_{i_1 i_2..i_{2n}}(t) 
=\bar P_{i_1 i_2..i_{2n}}(t) 
= 0$ for $n \geq 2$. 
Application of $\hat T_{ij}$ to the semi-classical state leads to
\bqa
\hat T_{ij} \ccr
& \approx &
\left[
2 \bar T_{ij} - 4 \bar T _{ik}  \bar P_{kl} \bar T_{lj} 
\right] \ccr
\eqa
to the leading order in $1/N$ and $\Delta$.
Here we use \eq{eq:TaT}
and 
$  \int D { \cT} ~ ( \sum_{b} \hat \phi_k^b \hat \phi_l^b ) ~\cb { \cT} \rb \chi({ \cT})
= \int D { \cT} ~ \cb { \cT} \rb \left( -  i  \frac{\partial}{\partial t_{kl}}  \right)  \chi({ \cT})$.
The latter relation implies that $\sum_{b} \hat \phi_k^b \hat \phi_l^b \sim N \bar P_{kl}$
plays the role of the conjugate momentum.
Therefore, 
$\hat T_{ij}$ acts as a classical variable 
in the space of semi-classical states.
This is true for more general semi-classical states 
in which higher order multi-local fields are turned on. 
%
%
%
%
%
%
%
%
%
If $\bar T _{ij}$ and $\bar P_{ij}$ decay in a distance function $d_{ij}$
associated with a lattice, the state exhibits a local structure. 
In this case, $\hat T_{ij} \ccr$ also decays exponentially following the local structure of the state,
and \eq{Hhat}  is well approximated by a local Hamiltonian.
For example,
only one-dimensional local hopping terms survive to the leading order in $1/N$
if the Hamiltonian is applied to $\ccr$ 
with $\hat T_{ij} \ccr  = \left[  e^{- [i-j]_L / \xi } + O(1/N) \right] \ccr$.

Now we examine time evolution of states in $\cW$
generated by $\hat H$.
An application of $ e^{-i dt \hat H}$ to $\cTr$ leads to
\bqa
&& e^{-i dt \hat H}  \cTr =   
 \int D \phi ~  
e^{-i dt N {\cal H}_0 \left[ {\cal T}, {\cal O} \right] } ~
e^{- N \sum_n \TTT \ppp }  
\cphir,
\label{eq:etHT}
\eqa
where ${\cal H}_0[ {\cal T}, {\cal O} ] $ is the induced Hamiltonian,
\bqa
&& {\cal H}_0[ {\cal T}, {\cal O} ]  = \nn 
&& -R'  \Bigl[ \sum_n 2n (2n-1) T_{ij j_1 j_2 .. j_{2n-2}} O_{ij} O_{j_1 j_2.. j_{2n-2}}   
 - \sum_{n,m} 4nm T_{j j_1j_2 .. j_{2n-1}} T_{i i_1 i_2 ..i_{2m-1}} O_{ij} O_{j_1 j_2 ..j_{2n-1} i_1i_2 .. i_{2m-1}} 
 \Bigr]    \nn
&&+ U
 \Bigl[  \sum_n 2n (2n-1) T_{ii j_1 j_2 .. j_{2n-2}} O_{j_1j_2 .. j_{2n-2}} 
-  \sum_{n,m} 4 n m T_{i j_1 j_2.. j_{2n-1}} T_{i i_1 i_2 ..j_{2m-1}} O_{j_1 j_2 ..j_{2n-1} i_1 i_2 .. i_{2m-1}} 
 \Bigr] + \lambda O_{ii}^2
 \nn
&&
+  \frac{R'}{N}
\Bigl[
\sum_n 2n(2n-1)  T_{ij j_1 j_2  .. j_{2n-2}} O_{ij  j_1j_2 .. j_{2n-2}} 
- \sum_{n,m} 4nm T_{j j_1 j_2.. j_{2n-1}} T_{i i_1 i_2 ..i_{2m-1}} O_{ij j_1 j_2 ..j_{2n-1} i_1 i_2 .. i_{2m-1}} 
\Bigr]
\Biggr\} \nn
\label{HTO}
\eqa
with $R'=\frac{N}{N-1}R$.
Because $\{ \cTr \} $ is a complete basis of $\cW$,
\eq{eq:etHT} can be written as a linear superposition
of $\cTr$,
\bqa
e^{-i dt \hat H} \ccr 
& = & \int  D {\cal T}^{(1)} D {\cal P}^{(1)} D {\cal T}^{(0)}
~ \cb \cT^{(1)} \rb ~
e^{ N \sum_n  \PPP^{(1)} \left( \TTT^{(1)} -  \TTT^{(0)} \right) 
- i N dt {\cal H}_0[ {\cal T}^{(0)}, {\cal P}^{(1)} ]  }
~ \chi(\cT^{(0)}), \nn
\label{ieH}
\eqa
where $ {\cal H}_0[ {\cal T}, {\cal P} ] $ is given by \eq{HTO} with 
${\cal O}$ replaced with ${\cal P}$.
The integration $T^{(1)}_{i_1 i_2 .. i_{2n}}$ in \eq{ieH} is defined
along the imaginary axis in the complex plane, 
where
 $\int D {\cal T}^{(1)} \equiv \int_{-\infty}^\infty 
 \prod_{i \geq j} dt_{ij}^{(1)}
 \prod_{i \geq j \geq k \geq l} dt_{ijkl}^{(1)} ..
 $ with $\TTT^{(1)} = i \ttt^{(1)}$.
 On the contrary, the integration over ${\cal P}$ is defined along the real axis.
 One can readily see that the integration over $ \ttt^{(1)}$ gives rise to the delta function,
 $\delta \left( 
 P^{(1)}_{i_1 i_2.. i_{2n}}  - O_{i_1 i_2.. i_{2n}}  
  \right)$.
 The following integration of ${\cal P}$ reproduces 
 \eq{eq:etHT}.

A finite time evolution is given by a path integration over the collective variables
and their conjugate momenta,
\bqa
e^{-i t \hat H} \ccr 
& = & \int  {\cal D} \cT {\cal D} \cP
~ \cb \cT(t) \rb ~
e^{ i S[ \cT, \cP ]  }
~ \chi(\cT(0)),
\label{eq:pi}
\eqa
where 
\bqa
S[ \cT, \cP ] &  = &  N \int_0^t d\tau 
\Bigl[
-i \sum_n \PPP \partial_\tau \TTT - {\cal H}_0[ \cT, \cP] 
\Bigr], 
\label{eq:STP}
\eqa
and 
$ {\cal D} \cT {\cal D} \cP  \equiv  \prod_{s=0}^{M_t} D \cT^{(s)} 
\prod_{s=1}^{M_t}D \cP^{(s)}$
with 
$M_t=t /dt$ and
$\tau =s ~ dt $.
The factor of $-i$ in \eq{eq:STP} is the reminder
that $\TTT$ should be integrated along the imaginary axes in the path integration.
It is noted that $-i \TTT$ and $\PPP$
are conjugate to each other.
Although ${\cal H}_0$ is not Hermitian,
it is $PT$-symmetric under which
$\ttt \rightarrow - \ttt$,
$\PPP \rightarrow \PPP$\cite{doi:10.1063/1.532860,Lee2016}. 
${\cal H}_0$  also has a real spectrum
because the original Hamiltonian in \eq{Hhat} is Hermitian.
${\cal H}_0$ can be mapped to a Hermitian Hamiltonian
through a similarity transformation.
However, we will proceed with the current representation of the Hamiltonian.

\section{ State dependent spread of entanglement }

\subsection{ Perturbative analysis near ultra-local solution }

 \begin{figure}[ht]
 \begin{center}
 \includegraphics[scale=0.7]{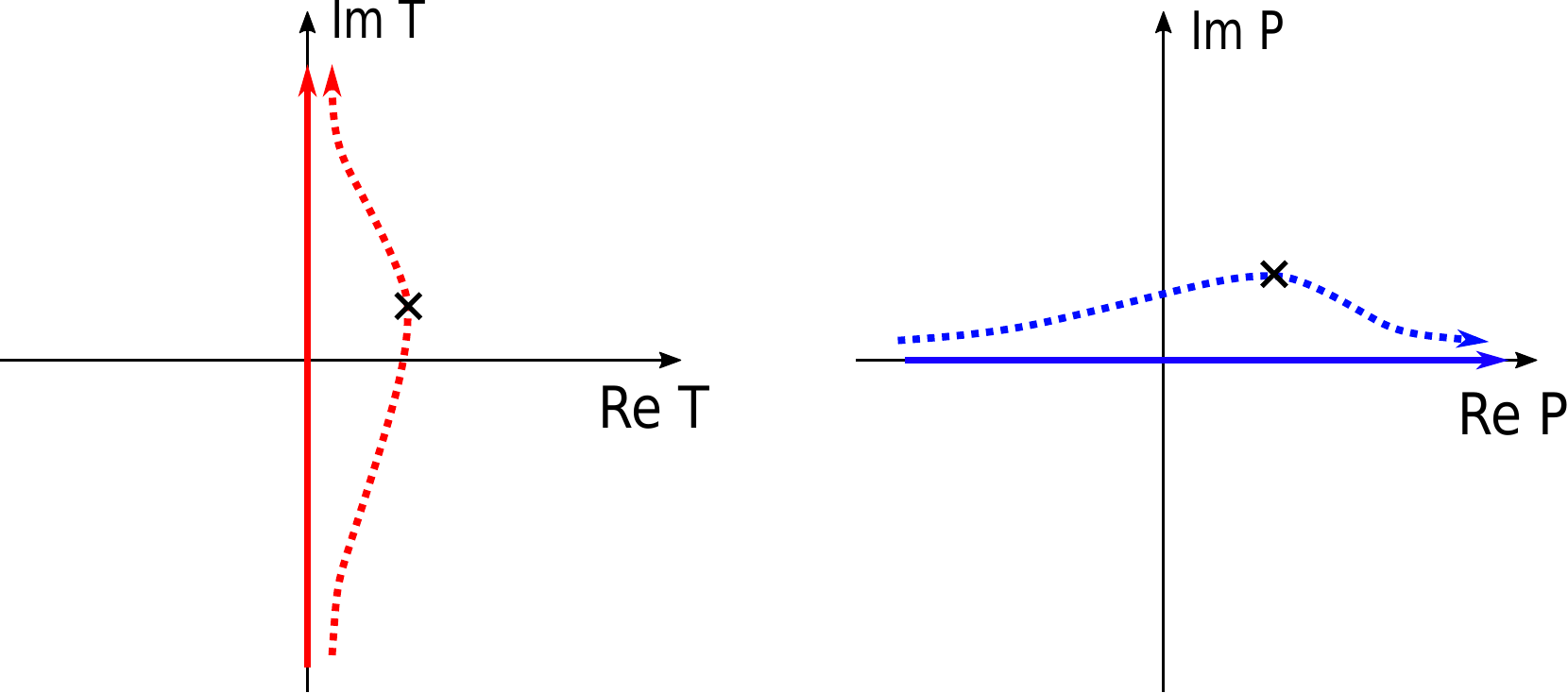} 
 \end{center}
 \caption{
 The contours of the path integrations in the complex planes of $\TTT$ and $\PPP$.
 While the original paths (solid lines) are defined along the imaginary and real axes respectively,
 the saddle points in general lie away from the original paths.
The saddle point can be reached by deforming the paths as denoted in the dashed lines.
Fluctuations around the saddle point are defined along the deformed paths. 
 }
 \label{fig:contour}
 \end{figure}

In this section, we examine time evolution of semi-classical states in \eq{cct}.
In the large $N$ limit,
the path integration in \eq{eq:pi} is dominated by the saddle point path
which satisfies the classical equation of motion.
While the integrations for $\TTT$ and $\PPP$ in \eq{eq:pi} are defined 
along the imginary and the real axes respectively,
they take general complex values at a saddle point.
This is illustrated in \fig{fig:contour}.

For general initial states in $\cW$, 
one should include all multi-local collective variables in the saddle-point solution.
Here we consider initial states 
with $ \bar T_{i_1 i_2..i_{2n}} = 0$
and $ \bar P_{i_1 i_2..i_{2n}} = 0$
for $n \geq 2$.
When only bi-local collective fields are turned on at $t=0$, 
$T_{i_1 i_2..i_{2n}}(t) = 0$ at all $t$ for $n \geq 2$ 
to the leading order in $1/N$.
This is due to the fact that
the initial state in $ \cV $ has the $O(N)$ symmetry, and 
the Hamiltonian in \eq{Hhat} also respects the $O(N)$ symmetry to the leading order in $1/N$.
Multi-local fields break the $O(N)$ symmetry,
and they are not generated to the leading order in $1/N$.
With $T_{i_1 i_2..i_{2n}}(t) = 0$ for $n \geq 2$,
the equations of motion for $T_{ij}$ and $P_{ij}$ 
decouple from $P_{i_1 i_2..i_{2n}}$ with $n \geq 2$ as well.
This allows one to use the effective Hamiltonian for the bi-local fields only,
\bqa
{\cal H}[ T, P ] 
& = &
R \left( - 2 T_{ij} P_{ji} + 4 T_{ik} P_{kl} T_{lj} P_{ji}   \right)   
+ U \left( 2 T_{ii} -  4 T_{ik} P_{kl}  T_{li}  \right) 
+ \lambda  P_{ii}^2
+ O\left(
N^{-1}
\right),
\label{eq:onH}
\eqa
which is obtained by turning off all multi-local fields with $n \geq 2$
in \eq{HTO}.
The bi-local fields obey the equation of motion,
\bqa
-i~ \partial_\tau T_{ij} & = &
R \left( - 2 T_{ij} + 8 T_{ik} P_{kl} T_{lj}  \right)   
- 4 U T_{ik} T_{kj}  
+ 2 \lambda  P_{ii}  \delta_{ij}, \nn
i~ \partial_\tau P_{ij} & = &
R \left( - 2 P_{ij} + 8 P_{ik} T_{kl} P_{lj}  \right)   
+ U \left( 
 2 \delta_{ij}
 - 4 P_{ik} T_{kj}
 - 4 T_{ik} P_{kj}
 \right)
\label{fulleom}
\eqa
with the initial condition
$T_{ij}(0) = \bar T_{ij}$,
$P_{ij}(0) = \bar P_{ij}$.

The spectrum of \eq{Hhat} is  bounded neither from above nor from below. 
However, the time evolution of a closed system with fixed energy is well defined
because the energy is conserved.
The same is true for the general relativity which has an unbounded spectrum.
The full Hilbert space is vast, and most states do not have a local structure.
Here we focus on a small fraction of initial states 
which exhibit local structures.
We do not attempt to find a dynamical mechanism
that may select such initial states.
Instead, our focus is to consider the time evolution of states with local structures,
and show that the relatively local Hamiltonian
acts as a local theory within the states with local structures.

Let us first consider ultra-local solutions.
For $T_{ij} = T_i \delta_{ij}$, $P_{ij} = P_i \delta_{ij}$, 
the equation of motion 
for each site is decoupled from others,
\bqa
-i~ \partial_\tau T_i & = & 
R \left( - 2 T_i + 8 T_i^2 P_i  \right)   
- 4 U T_i^2  
+ 2 \lambda  P_i, \nn
i ~\partial_\tau P_i & = & 
R \left( - 2 P_i + 8 P_i^2 T_i  \right)   
+ U \left( 
 2 - 8 P_i T_i
 \right).
\label{eq:ul}
\eqa
For $8 U^2 \lambda > R^3$,
there exists one ultra-local static solution (fixed point)
on the real axes of $T_i$ and $P_i$ :
$(T_*, P_* ) = 
\left(
\frac{1}{2} \left( \frac{\lambda}{U} \right)^{1/3},
\frac{1}{2} \left( \frac{U}{\lambda} \right)^{1/3}
\right)$.
For $8 U^2 \lambda \le R^3$,
two more fixed points coalesce on the real axes :
$(T_{**}^+, P_{**}^+ ) = 
\left(
\frac{R^2 + \sqrt{ R(R^3-8U^2 \lambda)}}{4UR} ,
\frac{U}{R} 
\right)$
and
$(T_{**}^-, P_{**}^- ) = 
\left(
\frac{R^2 - \sqrt{ R(R^3-8U^2 \lambda)}}{4UR} ,
\frac{U}{R} 
\right)$.
The static ultra-local solutions describe direct product states.
The direct product states remain to be direct product states
under the time evolution because the Hamiltonian 
acts as a ultra-local Hamiltonian
within the subspace of states 
without spatial entanglement.

 \begin{figure}[ht]
 \begin{center}
   \subfigure[]{
 \includegraphics[scale=0.5]{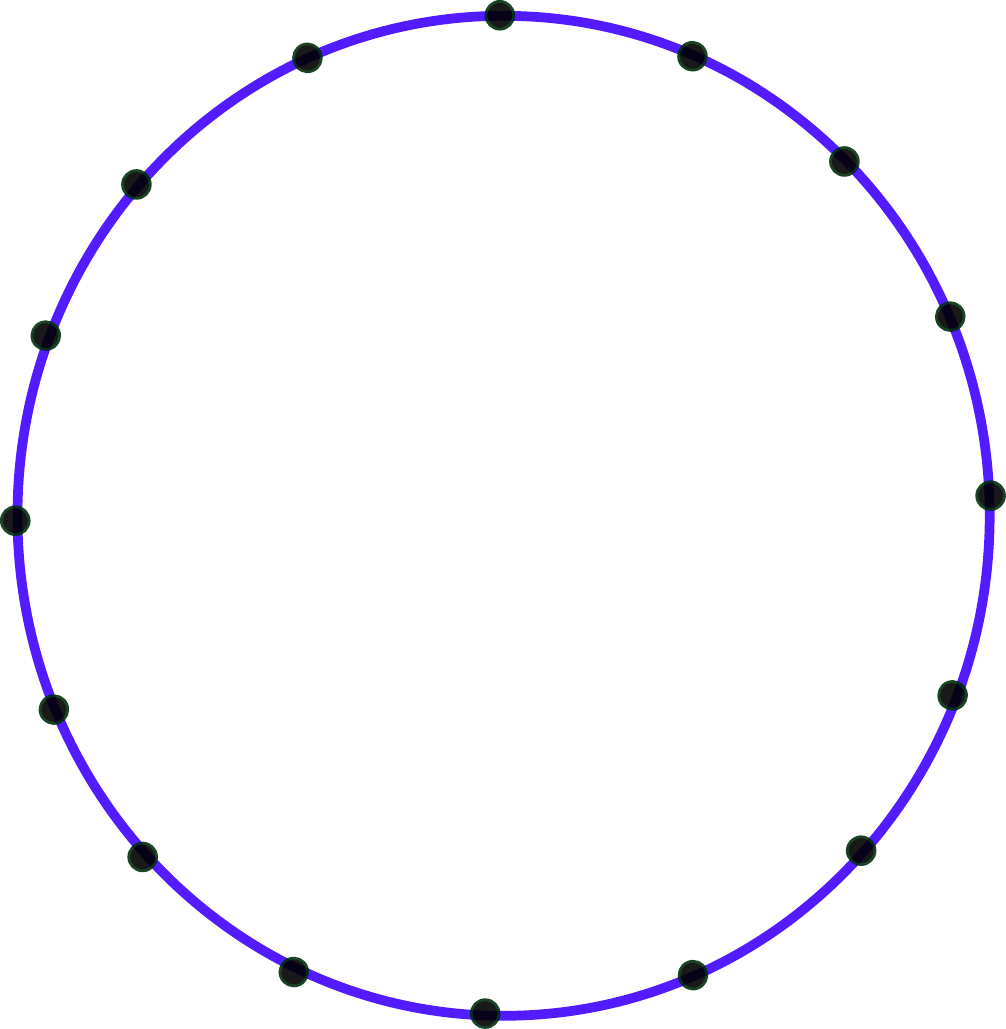} 
 } 
 \hfill
 \subfigure[]{
 \includegraphics[scale=0.5]{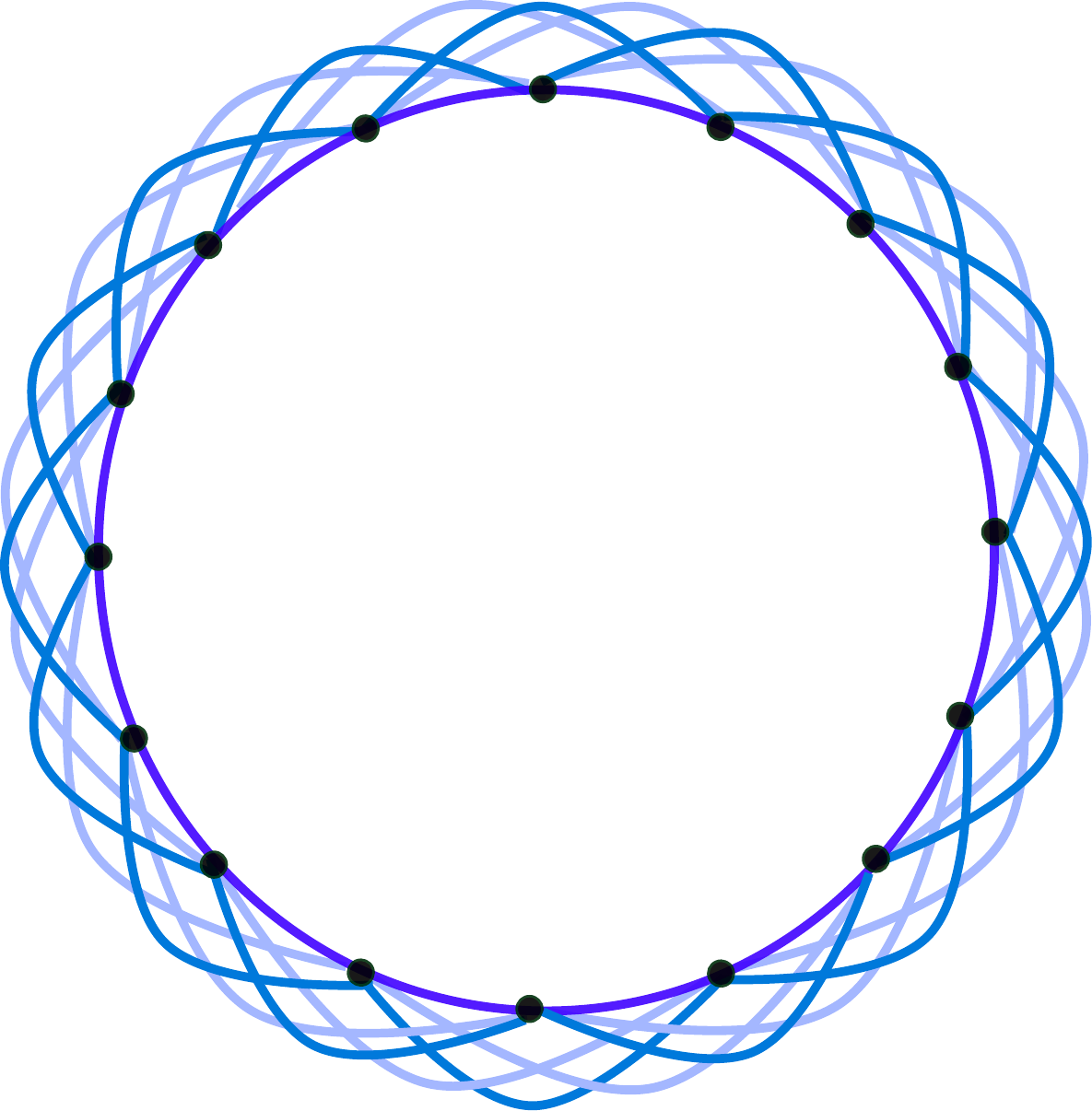} 
 } 
 \end{center}
 \caption{
 (a) The initial state in \eq{eq:initial} has only nearest neighbor entanglement bonds.
 (b) Under time evolution, further neighbor bonds are generated.
  }
 \label{fig:rings}
 \end{figure}

Here we focus on the case with $8 U^2 \lambda > R^3$,
and examine how initial states which are slightly perturbed away from 
the ultra-local fixed point evolve in time.
To be concrete, we consider the initial condition,
\bqa
T_{ij}(0) &=& T_* \delta_{ij}  + \epsilon \delta_{ d_{ij}^{(1)}, 1}, \nn
P_{ij}(0) &= & P_* \delta_{ij},
\label{eq:initial}
\eqa
where $d_{ij}^{(1)} =  [ i-j ]_L$.
The initial condition describes the ultra-local state 
perturbed with nearest neighbour entanglement bonds with strength $\epsilon$
that forms the one-dimensional lattice with the periodic boundary condition.
With the one-dimensional local structure, 
it is natural to use the label for each site as a coordinate of the lattice.
In the small $\epsilon$ limit, 
the initial state obeys the area law of entanglement :
the entanglement entropy of a sub-region 
scales as $-\epsilon^2 \ln \epsilon^2$, and  
is independent of the sub-system size to the leading order.
The solution in $t>0$ is written as
\bqa
T_{ij}(t) & = & T_{0,ij} + \epsilon ~ T_{1,ij}(t) + \epsilon^2 ~ T_{2,ij}(t) + ...,  \nn
P_{ij}(t) & = & P_{0,ij} + \epsilon ~ P_{1,ij}(t) + \epsilon^2 ~ P_{2,ij}(t) + ...~~  .
\label{expansion}
\eqa
Here $T_{0,ij} = T_* \delta_{ij}$
and
$P_{0,ij} = P_* \delta_{ij}$ denote
the unperturbed ultra-local static solution.
\eq{expansion} includes higher order terms in $\epsilon$ as well
because the non-linearity of the equation of motion generates
higher order terms at later time
even though the initial condition has only $O(\epsilon)$ 
perturbation.
Under time evolution, 
the second neighbor and further neighbor bi-local fields
are generated, creating longer-range entanglement in the system.
This is illustrated in \fig{fig:rings}.
The increase in the range of the bi-local fields in real space  
describes how entanglement spreads in space.
For earlier studies of spread of entanglement in local and non-local Hamiltonians,
see Refs. 
\cite{
Nozaki2013,
Jurcevic:2014aa,
Rangamani2016,
Casini2016,
PhysRevLett.112.011601,
Mezei2017,
PhysRevD.95.086008}.

In organizing the equation of motion 
as a perturbative series in $\epsilon$,
it is convenient to combine the phase space variables at each order into a two-component complex vector,
$\vec V_{n,ij} = \left( \begin{array}{c} T_{n,ij} \\ P_{n,ij} \end{array} \right)$.
The phase space vectors obey the equations of motion,
\bqa
\partial_t \vec V_{n,ij} & = & M_{ij} \vec V_{n,ij} + \vec A_{n,ij},
\label{eq:V}
\eqa
where
\bqa
M_{ij} & = & i \left(
\begin{array}{cc} 
-2R + 16R T_* P_* - 8U T_* &   8R T_*^{2} + 2 \lambda \delta_{ij} \\
- 8R P_*^{2} + 8U P_* & 
2R - 16R T_* P_* + 8U T_*
\label{eq:M}
\end{array}
\right),
\eqa
and 
\bqa
\vec A_{1,ij} & = & 0, \nn
\vec A_{n,ij} &= & i \left( \begin{array}{c}
8R  \sum\limits_{ a,b,c < n} T_{a,ik} P_{b,kl} T_{c,lj} ~\delta_{a+b+c,n}
- 4U \sum\limits_{a,b<n} T_{a,ik} T_{b,kj}~ \delta_{a+b,n} \\
-8R \sum\limits_{a,b,c < n} P_{a,ik} T_{b,kl} P_{c,lj}~  \delta_{a+b+c,n}
+ 4U \sum\limits_{a,b<n} 
(
P_{a,ik} T_{b,kj} 
+ T_{a,ik} P_{b,kj} 
)~ \delta_{a+b,n}
\end{array} \right)
\label{eq:A}
\eqa
for $n>1$.
In Eqs. (\ref{eq:V}) - (\ref{eq:A}),  repeated site indices are summed over all sites 
except for $i,j$.
$M_{ij}$ is the linearized Hamiltonian that dictates 
the orbit of $\vec V_{n,ij}$ near the ultra-local solution.
For $i=j$, 
$M_{ij}$ has eigenvalues $\pm i \omega_0$ with 
$ \omega_0  \equiv  2  \sqrt{ 3 U^{2/3} \lambda^{1/3} ( 2 U^{2/3} \lambda^{1/3} - R ) }$,
and associated  
eigenvectors 
$\vec u_\pm^T
=\left(
\frac{ 
R U^{2/3} \lambda^{2/3} 
- 2  U^{4/3} \lambda 
\pm \sqrt{ 3 U^{2} \lambda^{5/3} ( 2 U^{2/3} \lambda^{1/3} - R ) }}
{  -R U^{4/3} + 2 U^2 \lambda^{1/3} } , 
1
\right)$.
For $i \neq j$,
$M_{ij}$ has eigenvalues 
$\pm i \omega_1$ with 
$ \omega_1 \equiv  \sqrt{ \frac{2}{3} } \omega_0$,
and 
eigenvectors 
$\vec v_\pm^T
=\left(
\frac{ 
R U^{2/3} \lambda^{2/3} 
- 2  U^{4/3} \lambda 
\pm \sqrt{ 2 U^{2} \lambda^{5/3} ( 2 U^{2/3} \lambda^{1/3} - R ) }}
{  -R U^{4/3} + 2 U^2 \lambda^{1/3} } , 
1
\right)$.
$\vec A_{n,ij}$ is the `force' term for $\vec V_{n,ij}$ generated from 
$\vec V_{m,kl}$ with $m < n$.

Now we examine the equations of motion order by order in $\epsilon$.
For $n=1$, there is no force term,
and the equation for $\vec V_{1,ij}$ is decoupled for each $i,j$. 
The solution to \eq{eq:V} is given by
\bqa
\vec V_{1,ij}(t)  &=&
\left[ \alpha_{1,ij} e^{i \omega_1 t}  \vec v_{+} + \beta_{1,ij} e^{-i \omega_1 t} \vec v_{-}  \right] \delta_{d_{ij},1},
\label{eq:v1}
\eqa
where $\alpha_{1,ij}, \beta_{1,ij}$ are constants determined from the initial condition,
$\vec V_{1,ij}^T(0) =  (  \delta_{d_{ij},1} ,0 )$.
$\vec V_{1,ij}(t)$ on the nearest neighbor bonds undergo independent oscillatory motions,
and $V_{1,ij}(t) = 0$ for $d_{ij} \neq 1$.

At the second order in $\epsilon$, 
$\vec V_{2,ij}$ is driven by the force term which is quadratic in $\vec V_{1,ij}$.
Because $A_{2,ij} \sim \sum_k V_{1,ik} V_{1,kj} $, 
$A_{2,ij}$ oscillates with frequencies
$C_2 = \{ 0, \pm 2 \omega_1  \}$
for $i,j$ with $d_{ij}=0, 2$.
The force term can be written as
$\vec A_{2,ij} = \sum_{ \Omega_p \in C_2 }
e^{ i \Omega_p t}
\vec a_{2,ij}^{p}  
$,
and the solution to the driven  oscillator becomes
\bqa
\vec V_{2,ij}(t) =
\left\{ \begin{array}{cc}
\alpha_{2,ii} e^{i \omega_{0} t}  \vec u_{+ } + \beta_{2,ii} e^{-i \omega_{0} t} \vec u_{-} + \sum_{\Omega_p \in C_2} e^{i \Omega_p t} (  i \Omega_p - M_{ii} )^{-1} \vec a_{2,ii}^{p} & ~~ \mbox{ for  $i=j$}, \\
 \alpha_{2,ij} e^{i \omega_{1} t}  \vec v_{+ } + \beta_{2,ij} e^{-i \omega_{1} t} \vec v_{-} + \sum_{\Omega_p \in C_2} e^{i \Omega_p t} (  i \Omega_p - M_{ij} )^{-1} \vec a_{2,ij}^{p} &~~ \mbox{ for $i,j$ with $d_{ij}=2$}, 
 \end{array} \right.
\label{eq:v2}
\eqa
where $\alpha_{2,ij}, \beta_{2,ij}$ are determined from $\vec V_{2,ij}(0)=0$.
For all other $i,j$ with $d_{ij}  \neq 0,2$, $V_{2,ij}(t)=0$.
It is noted that  
$( i \Omega_p - M_{ij})$ is invertible for all $\Omega_p$ in $C_2$ 
because the driving frequencies are not resonant with the natural frequency of $M_{ij}$.
At this order, the second nearest neighbour entanglement bonds are generated,
but their amplitudes remain to be $O(\epsilon^2)$.
This describes a non-propagating evanescent mode.

At the third order, orbits are no longer quasi-periodic.
Resonances occur because $\vec A_{3,ij}$ includes driving force 
with frequencies $\pm \omega_1$,
which is the natural frequency of $M_{ij}$. 
For example,  
$A_{3,i,i+1} \sim V_{1,i,i+1} V_{1,i+1,i} V_{1,i,i+1}$
and  
$A_{3,i,i+3} \sim V_{1,i,i+1} V_{1,i+1,i+2} V_{1,i+2,i+3}$  
generate forces with frequencies $\pm \omega_1$.
Besides oscillatory components,
the resonances lead to a linear growth of the amplitude as
 \bqa
\vec V_{3,ij}(t) &=&
 t \left( \alpha_{3,ij}  e^{i \omega_1 t} \vec v_{+}  + \beta_{3,ij}  e^{-i \omega_1 t} \vec v_{-} \right) + ... 
 ~~ \mbox{for $i,j$ with $d_{ij}=1,3$}.
\eqa
Here $...$ denotes pure oscillatory parts.
Through the resonance, the amplitudes of the third nearest neighbor bi-local fields increase.
This results in a growth of the range of the bi-local fields in real space as is shown 
in \fig{fig:rings}.
Since it takes $t \sim \epsilon^{-2}$
for the third nearest neighbor bi-local field to become $O(\epsilon)$,
the coordinate speed at which the range of the bi-local field grows in real space
vanishes as $O(\epsilon^2)$
in the small $\epsilon$ limit.

We emphasize the fact that further neighbor bi-local fields are generated only at higher orders,
and they are exponentially suppressed in $\epsilon$ at small $t$.
This is because the initial state provides only nearest neighbor entanglement bonds
that source the growth of further neighbor bi-local fields.
The locality of the theory that governs the evolution of the bi-local fields at early time 
is determined by the local structure of the initial state. 
At later time, the range of the bi-local fields becomes larger in coordinate distance,
and the local structure of the state evolves.
The emergent theory that governs the evolution of the bi-local fields at a given time
is local only at scales larger than the range of the bi-local fields at that moment.
Furthermore, the dimension of the emergent local theory is determined by the dimension
of the local structure of the initial state.
If one starts with an initial state with a two-dimensional local structure,
the entanglement spreads following the two-dimensional lattice set by the initial state.


\subsection{ Numerical solution }

 \begin{figure}[ht]
 \begin{center}
   \subfigure[]{
 \includegraphics[scale=0.45]{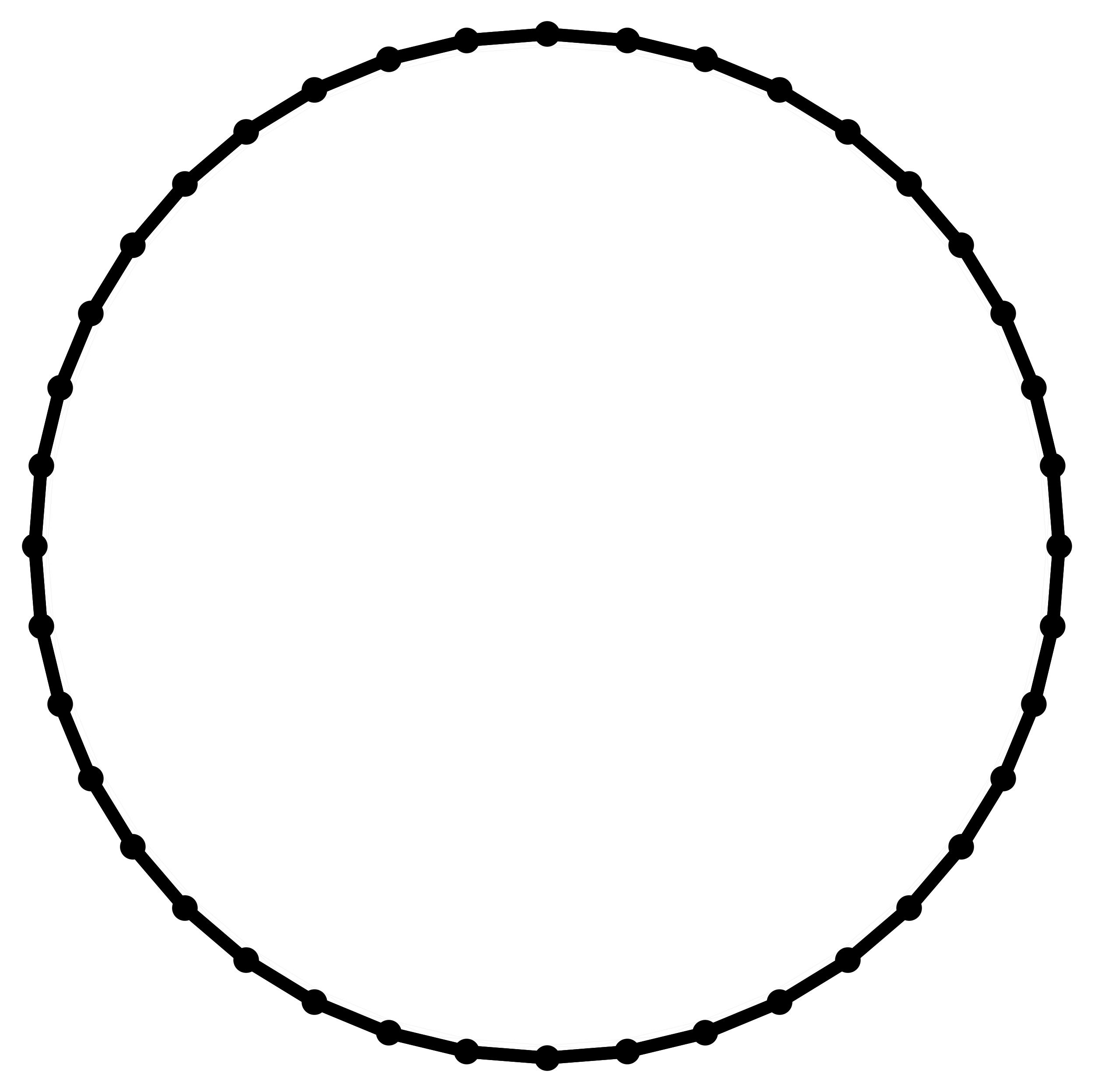} 
 } 
 \hfill
 \subfigure[]{
 \includegraphics[scale=0.45]{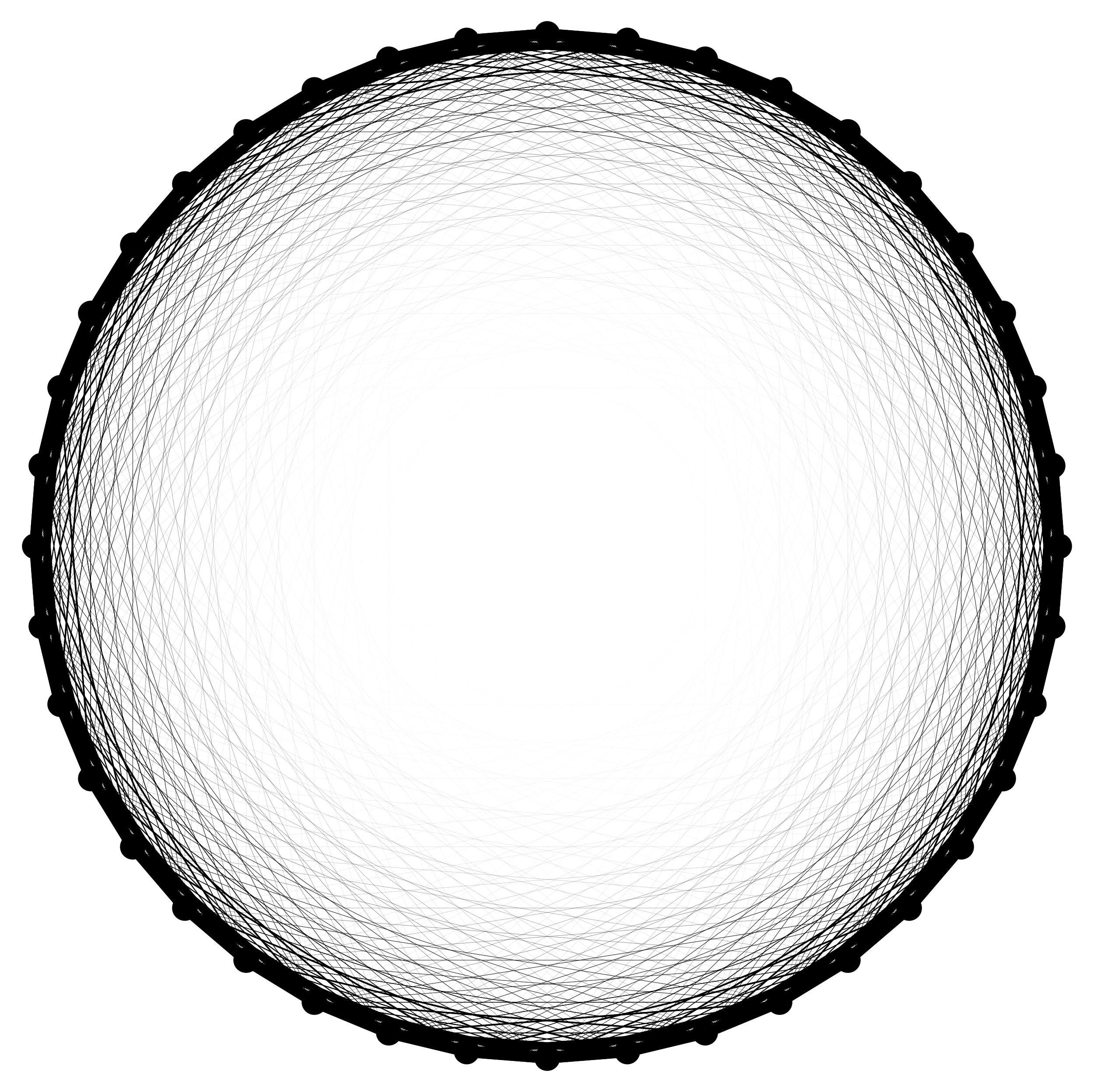} 
 } 
  \subfigure[]{
 \includegraphics[scale=0.45]{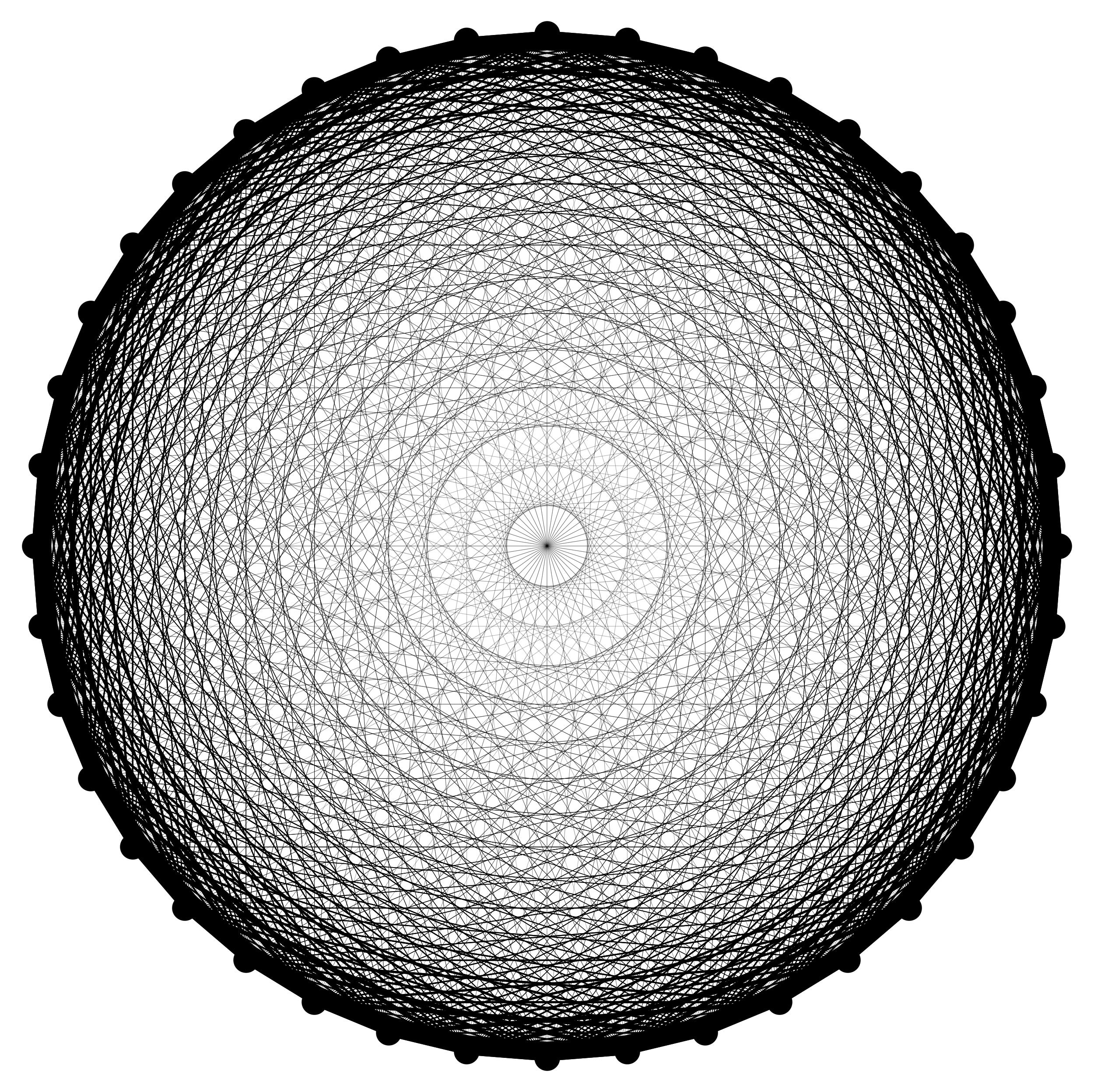} 
 } 
 \hfill
 \subfigure[]{
 \includegraphics[scale=0.45]{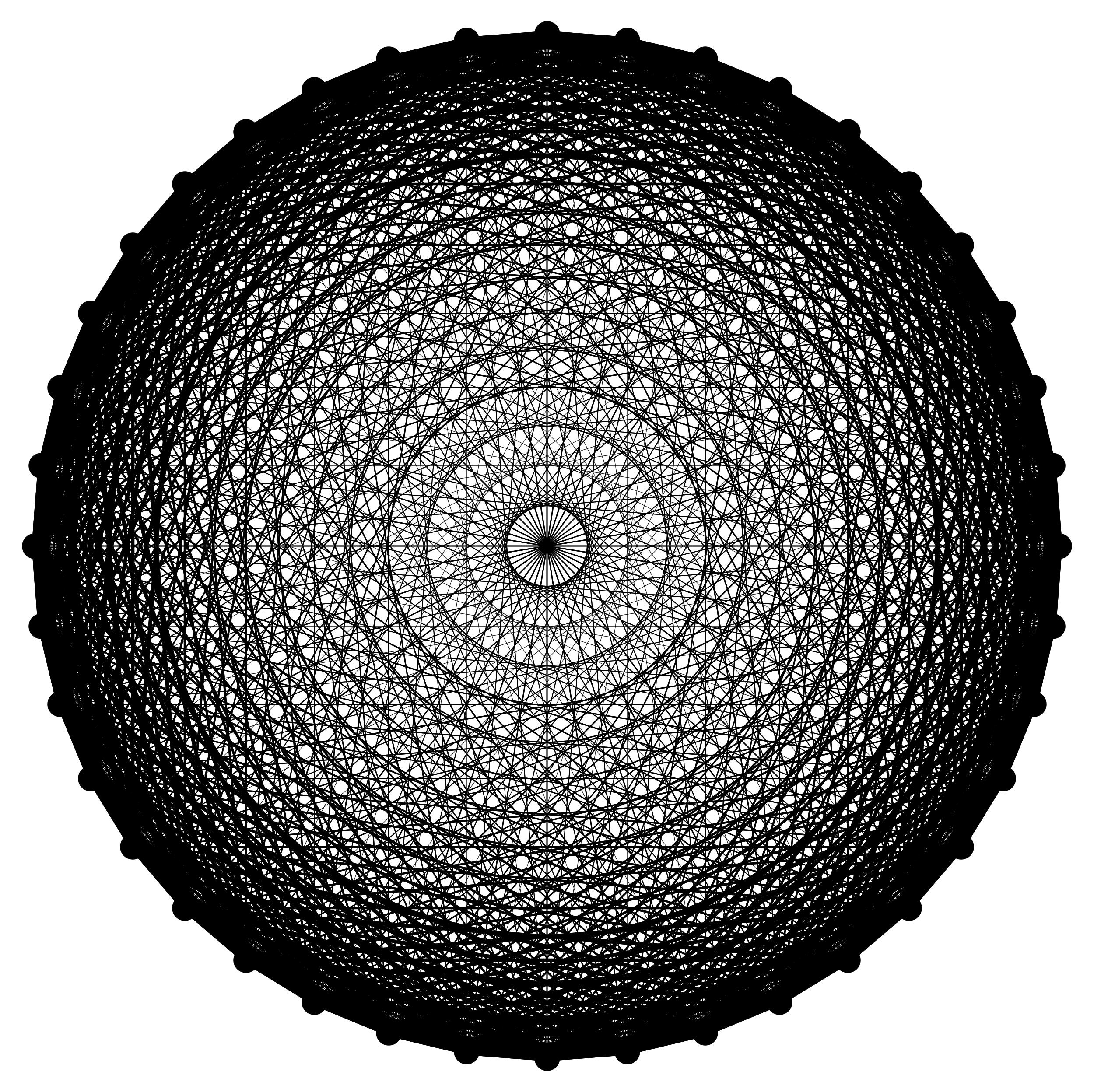} 
 } 
  \end{center}
 \caption{
Snapshots of the graphs that arise from
the numerical solution of $T_{ij}(t)$
at
 (a) $t=0$,
 (b) $t=15$,
 (c) $t=30$,
 (d) $t=45$
 for 
$\epsilon=0.12$ with
$R=U=\lambda=1$ and $L=40$.
Dots represent sites, and
the thickness of the line between sites $i$ and $j$ is  
proportional to $|T_{ij}|$. 
 }
 \label{fig:emergent_graph}
 \end{figure}

 \begin{figure}[ht]
 \begin{center}
   \subfigure[]{
 \includegraphics[scale=0.7]{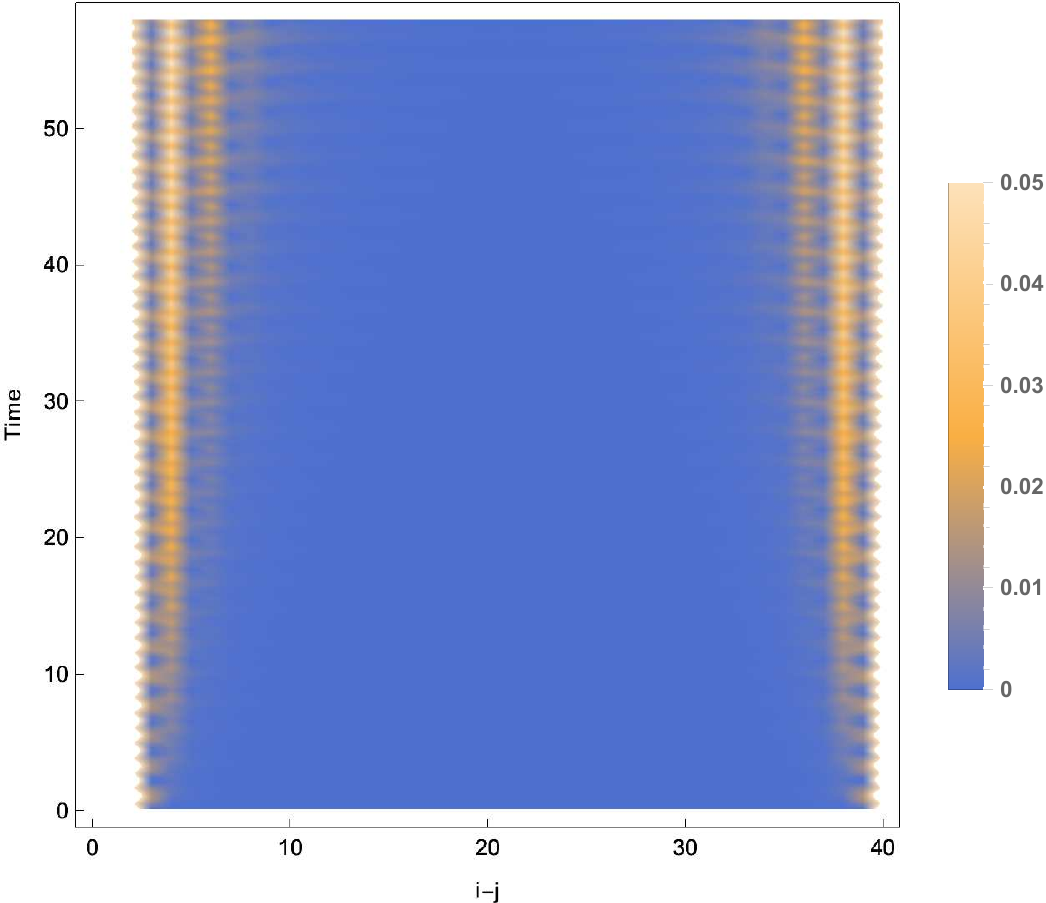} 
 } 
 \hfill
 \subfigure[]{
 \includegraphics[scale=0.7]{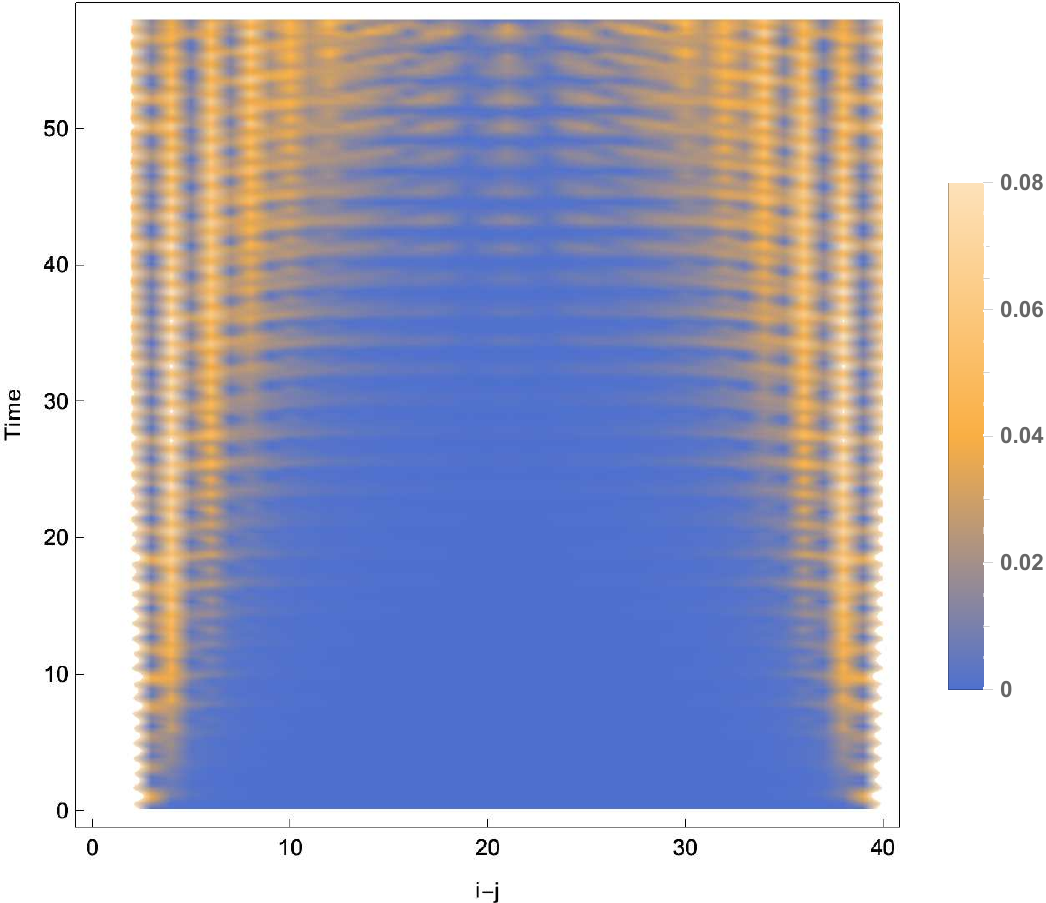} 
 } 
  \subfigure[]{
 \includegraphics[scale=0.7]{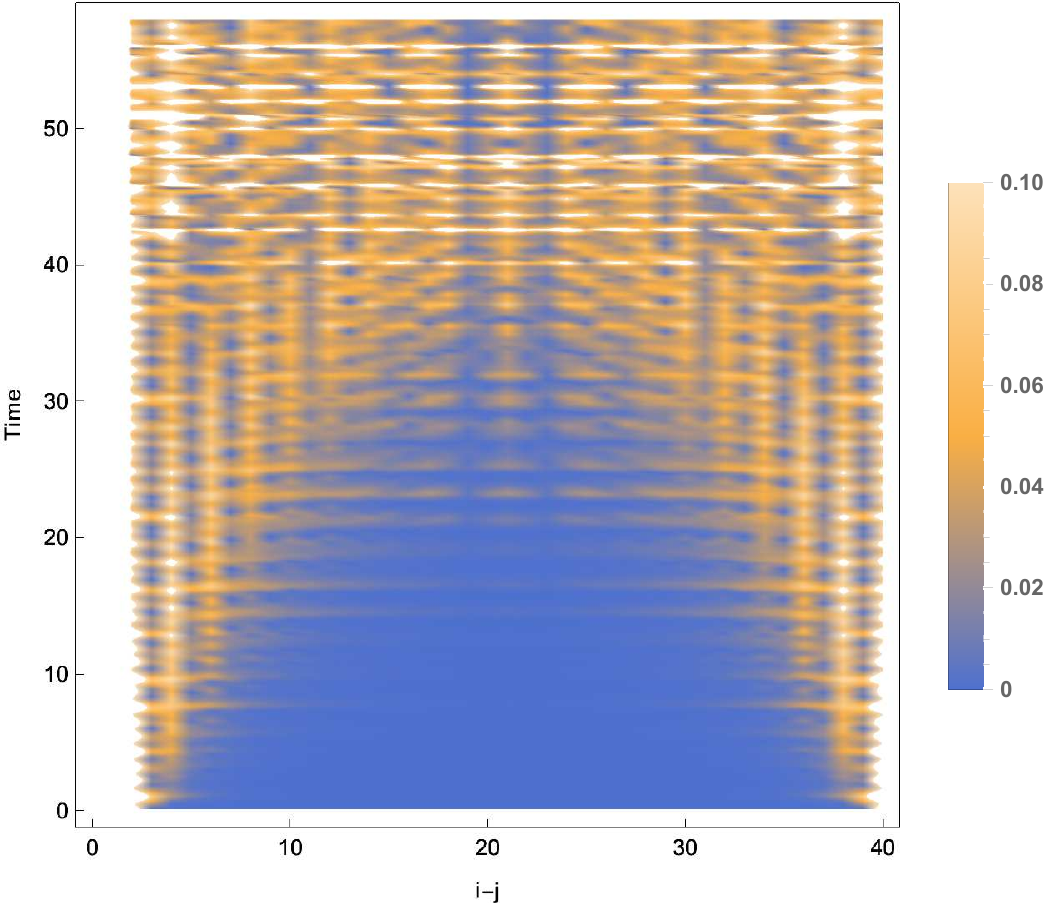} 
 } 
 \hfill
 \subfigure[]{
 \includegraphics[scale=0.7]{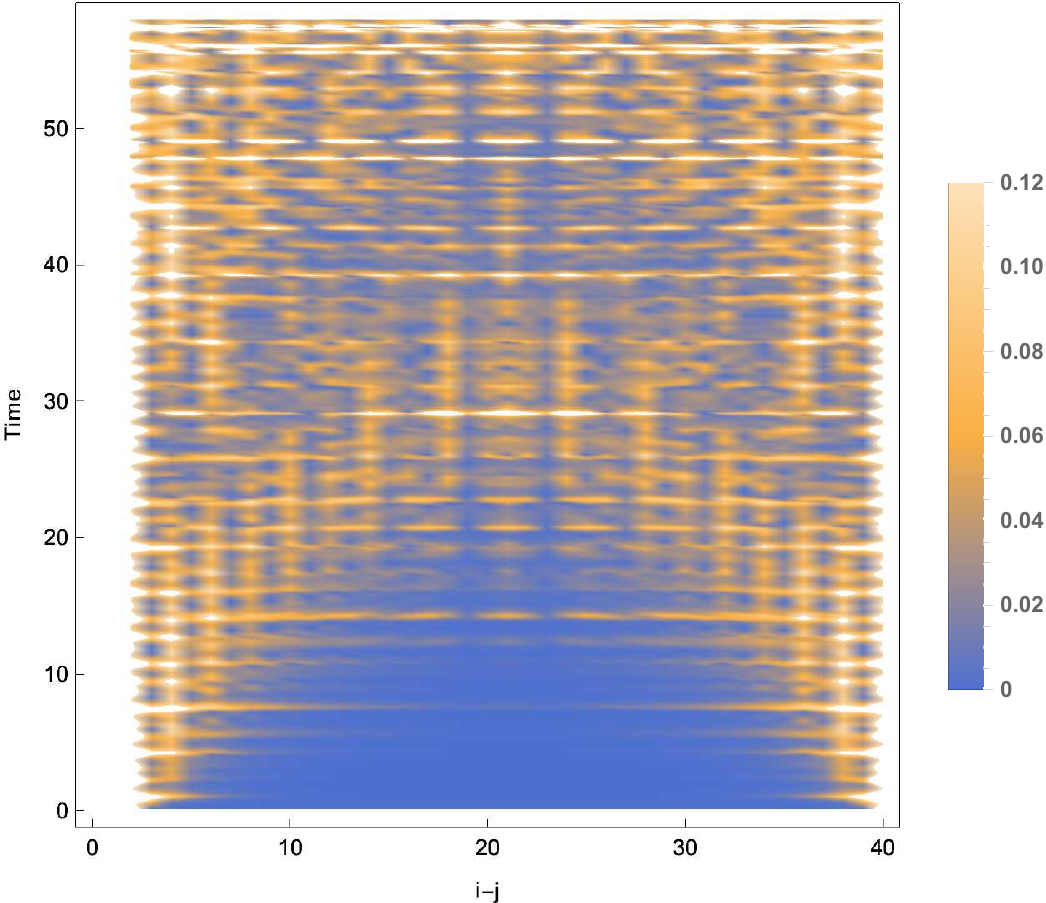} 
 } 
 \end{center}
 \caption{
Density plot of $|T_{ij}(t)|$ shown as a function of $i-j$ and $t$
for
 (a) $\epsilon=0.08$,
 (b) $\epsilon=0.12$,
 (c) $\epsilon=0.16$,
 (d) $\epsilon=0.2$
 with $R=U=\lambda=1$
 and $L=40$.
 }
 \label{fig:Tij3D}
 \end{figure}

In order to test the predictions of the perturbative analysis, 
we now solve \eq{fulleom} numerically for the initial condition in \eq{eq:initial}.
In \fig{fig:emergent_graph}, the graphs that emerge from $T_{ij}(t)$ are shown at different time slices for $\epsilon=0.12$,
where the thickness of the lines connecting sites are drawn in proportion to the magnitude of the bi-local field.
At small $t$, sites are connected mainly through short-range bonds, 
maintaining the one-dimensional local structure.
As time increases, the range of the bi-local fields grows,
and all sites get connected to all other sites at late time. 
\fig{fig:Tij3D} shows the evolution of $T_{ij}(t)$ as a function of 
$i-j$ and $t$ for various choices of $\epsilon$.
The translational symmetry of the initial condition guarantees that 
$T_{ij}(t)$ and $P_{ij}(t)$ depends on $i$ and $j$ only through $i-j$. 
Furthermore, $T_{ij}(t) = T_{i L+2i-j}(t)$ due to the periodic boundary condition.
The profile of $P_{ij}(t)$ 
in the space of $i-j$ and $t$ 
is similar to that of $T_{ij}(t)$
shown in \fig{fig:Tij3D}. 
As expected, 
the speed at which the range of the bi-local fields increases in real space
strongly depends on $\epsilon$.

 \begin{figure}[ht]
 \begin{center}
   \subfigure[]{
 \includegraphics[scale=0.5]{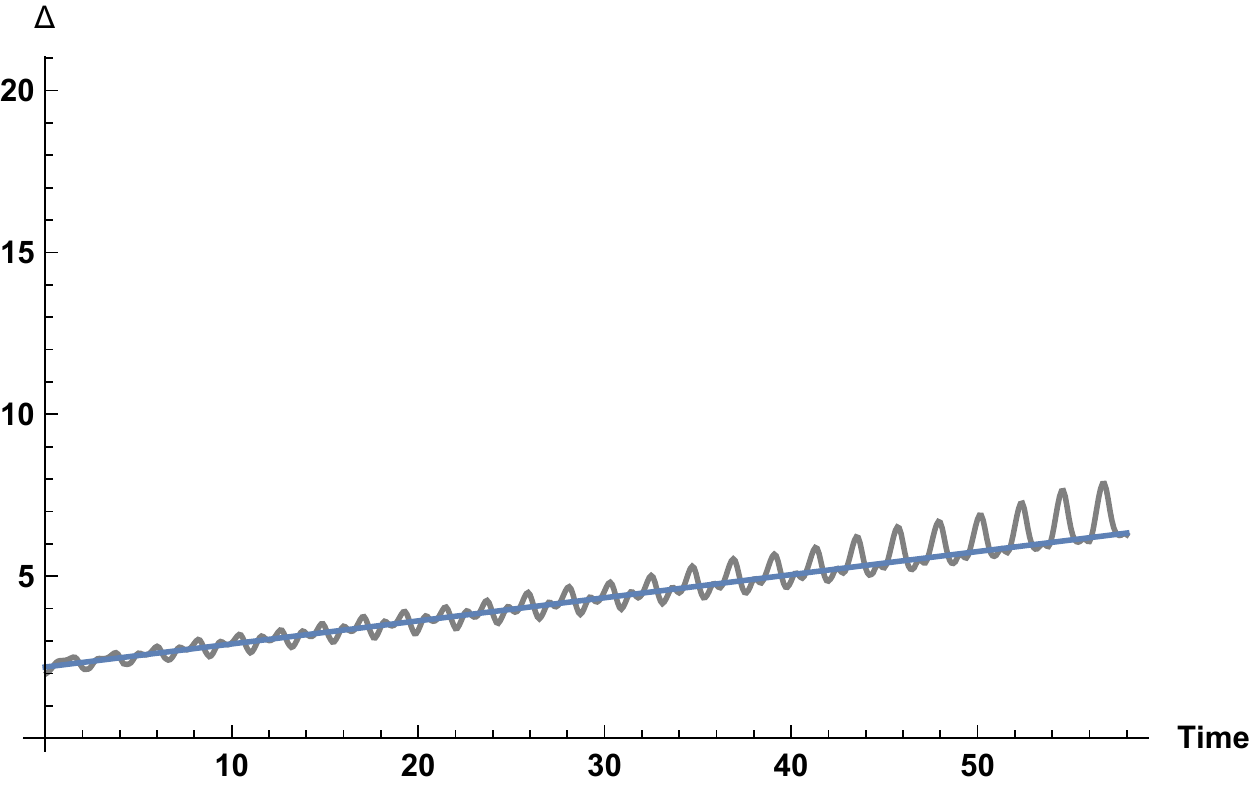} 
 } 
 \hfill
 \subfigure[]{
 \includegraphics[scale=0.5]{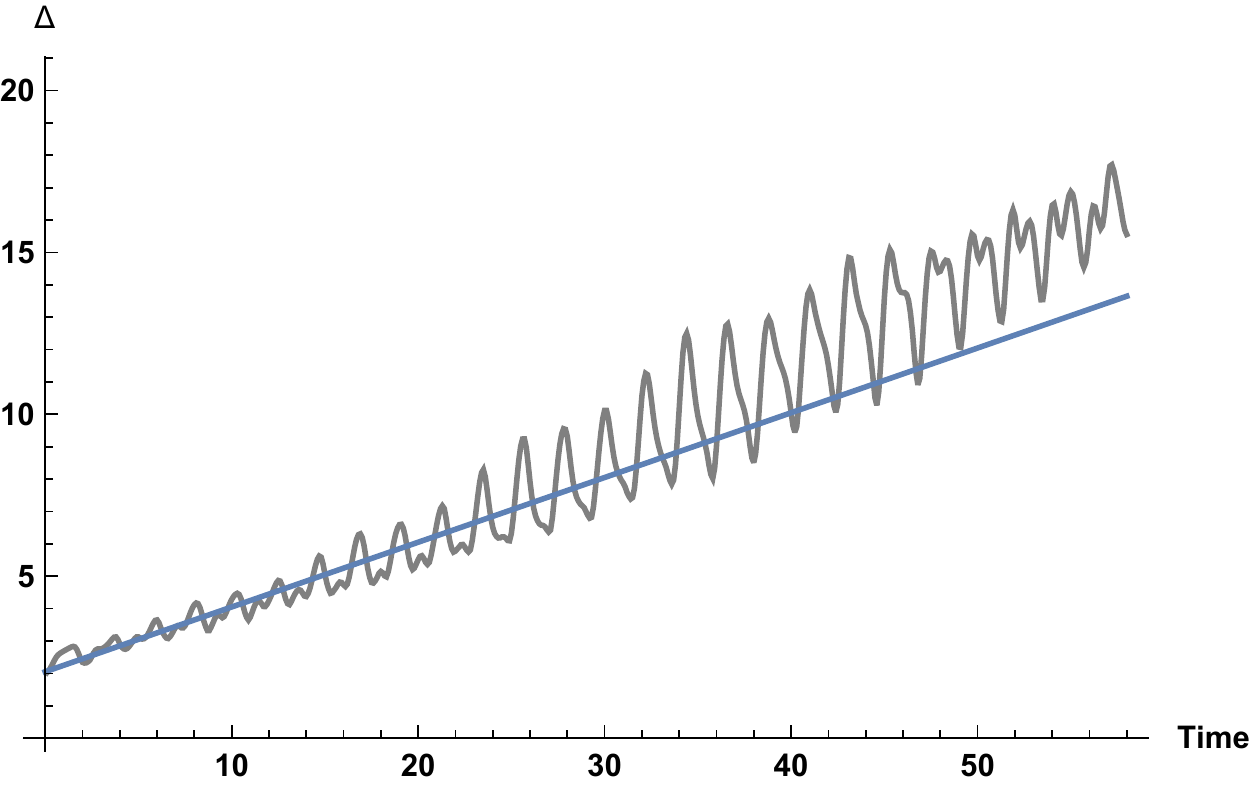} 
 } 
  \subfigure[]{
 \includegraphics[scale=0.5]{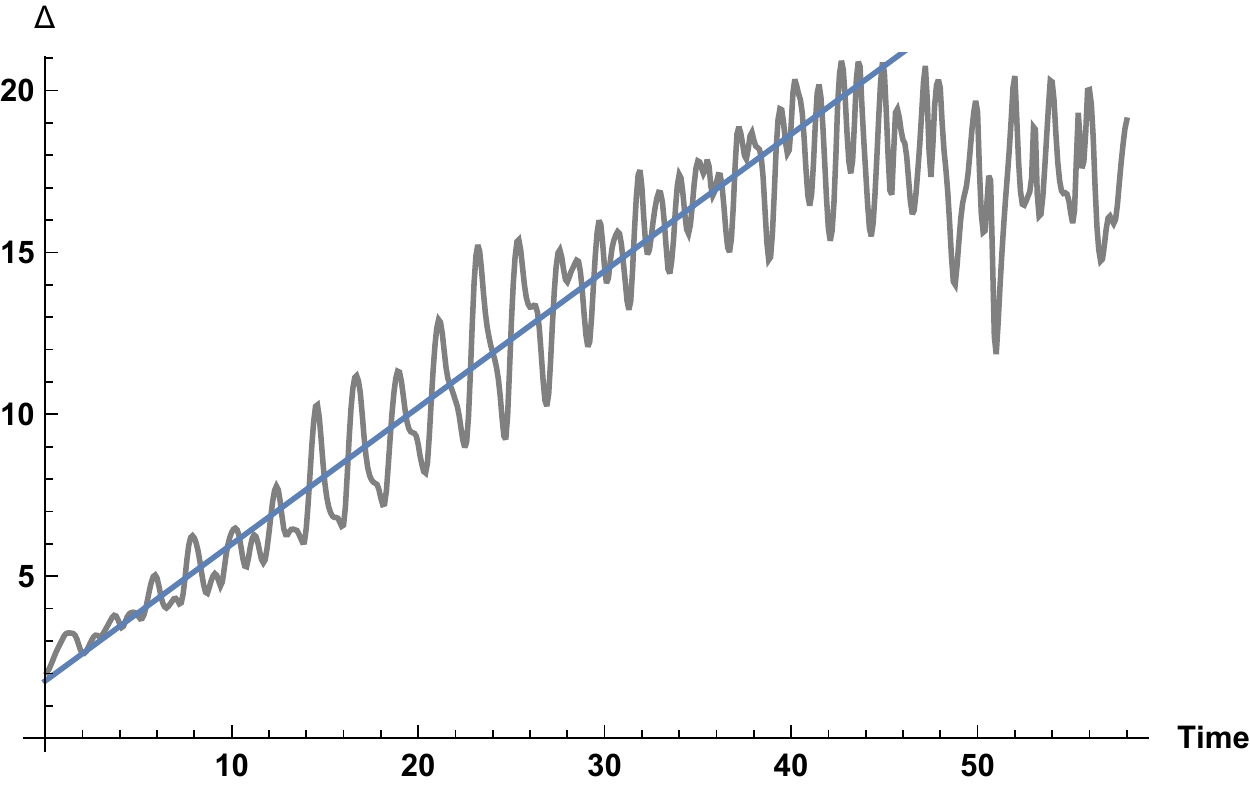} 
 } 
 \hfill
 \subfigure[]{
 \includegraphics[scale=0.5]{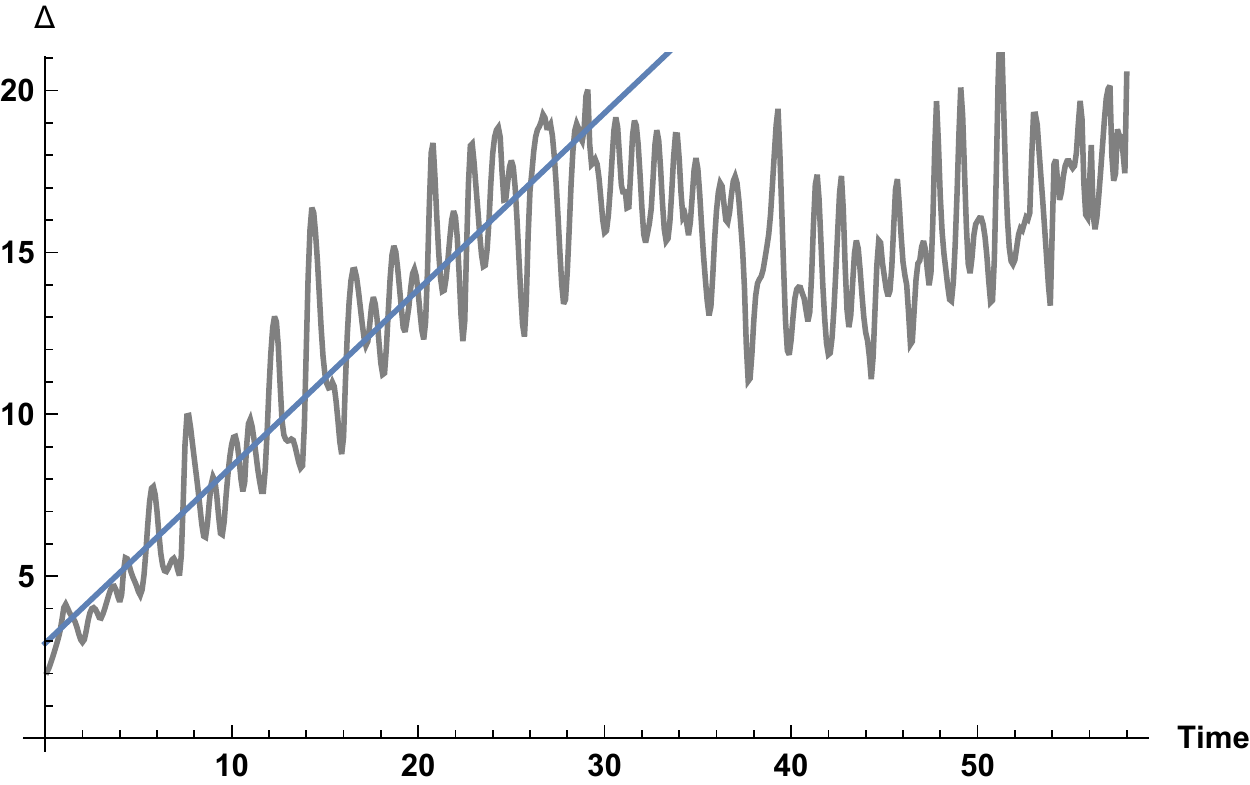} 
 } 
 \end{center}
 \caption{
The  range of the bi-local field defined in \eq{Delta} 
plotted as a function of time
for
 (a) $\epsilon=0.08$,
 (b) $\epsilon=0.12$,
 (c) $\epsilon=0.16$,
 (d) $\epsilon=0.2$
 with $R=U=\lambda=1$
 and $L=40$.
The straight lines represent the best linear fits of $\Delta(t)$ in $0 < t < 30$.
 }
 \label{fig:Tijsize}
 \end{figure}

 \begin{figure}[ht]
 \begin{center}
 \includegraphics[scale=0.7]{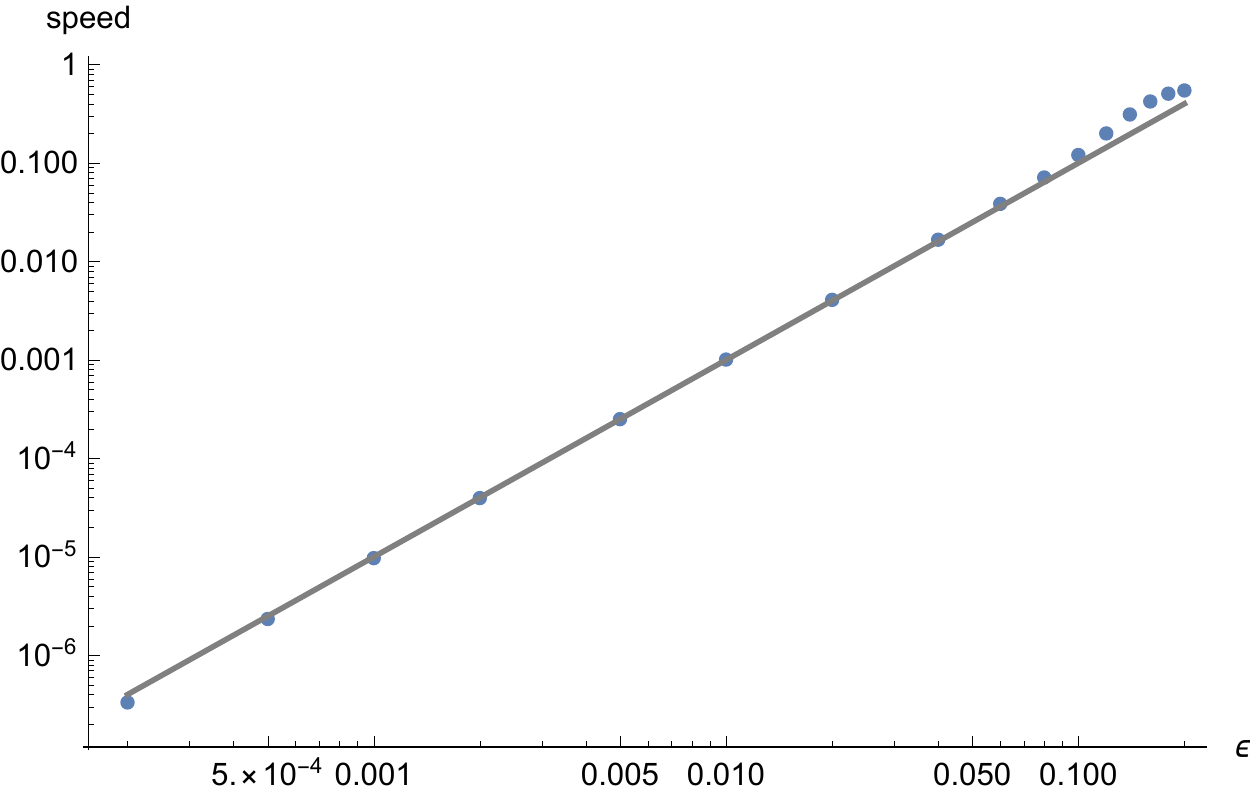} 
 \end{center}
 \caption{
The rate at which $\Delta(t)$ increases in time  
obtained from a linear fit of $\Delta(t)$ in $0 < t < 30$ 
plotted in the log-log scale.
The straight line is $10 \epsilon^2$.
 }
 \label{fig:speed}
 \end{figure}

In order to quantify the speed of entanglement spread, 
we introduce
\bqa
\Delta(t) \equiv 2 \frac{ \sum_{ 1 \leq |i-j|\leq L/2 } |T_{ij}(t)| |i-j| }{  \sum_{ 1 \leq |i-j|\leq L/2 } |T_{ij}(t)| }
\label{Delta}
\eqa
which measures the range of the bi-local field in coordinate distance. 
Here the maximum size of the bi-local field is restricted to $|i-j| \leq L/2$ 
because of the periodic boundary condition.
For the initial state with only short-range entanglement bonds, $\Delta(0) \sim O(1)$.
In the other extreme limit, 
if $T_{ij}(t)$ spreads over the entire system and becomes independent of $i-j$,
$\Delta(t)$ approaches  $L/2$.
In \fig{fig:Tijsize}, we show the growth of  $\Delta(t)$ for various values of $\epsilon$.
At small $t$, $\Delta(t)$ increases linearly modulated by oscillatory contribution.
The speed of the linear growth of $\Delta (t)$ at early time  
is plotted as a function of $\epsilon$ in \fig{fig:speed}.
In the small $\epsilon$ limit, the speed scales with $\epsilon^2$,
which is consistent with the perturbative analysis.
This strong state dependence of the speed is contrasted to 
local Hamiltonians which typically exhibit
an $O(1)$ dependence of speed on state\cite{PhysRevB.97.205103}.

At larger $t$, 
the growth of $\Delta(t)$ 
exhibits an acceleration in time
as is shown in \fig{fig:Tijsize}.
This non-linearity is more noticeable 
for small $\epsilon$ in which 
there is enough time before the range of the bi-local fields
reach the system size.
The acceleration can be attributed to the fact that 
further neighbor entanglement bonds can be created more easily 
once the state develops bonds beyond nearest neighbor sites in $t>0$. 
Eventually, $\Delta(t)$ stops growing 
once the range of the bi-local fields becomes comparable to the system size.
In the late time limit, the bi-local field spreads over the entire system
as is shown in Figs. \ref{fig:emergent_graph} and \ref{fig:Tij3D}.
This implies that the state in the large $t$ limit supports entanglement entropy that scales 
with the coordinate volume of sub-systems.
The time it takes for the system to reach 
such a state strongly depends on the amount of entanglement in the initial state.

State dependent coordinate speed can be understood as originating 
from state dependent geometry.
It is noted that $\Delta(t)$ determines the range of hopping at time $t$ in \eq{Hhat}.
If we define the {\it proper distance} such that there exist
hoppings only between sites within unit proper distance,
sites which are separated by $\Delta(t)$ in {\it coordinate distance}
are regarded to be within a unit proper distance.
As $\Delta(t)$ increases in time, 
the proper size of the system decreases.
In this sense, the time evolution shown in \fig{fig:Tij3D}
can be viewed as a `collapsing universe',  
where the proper size of the system shrinks in time
\footnote{
It is of interest to understand if the big crunch will be followed 
by a bounce of an expanding space at a later time.
}.

 \begin{figure}[ht]
 \begin{center}
 \includegraphics[scale=0.5]{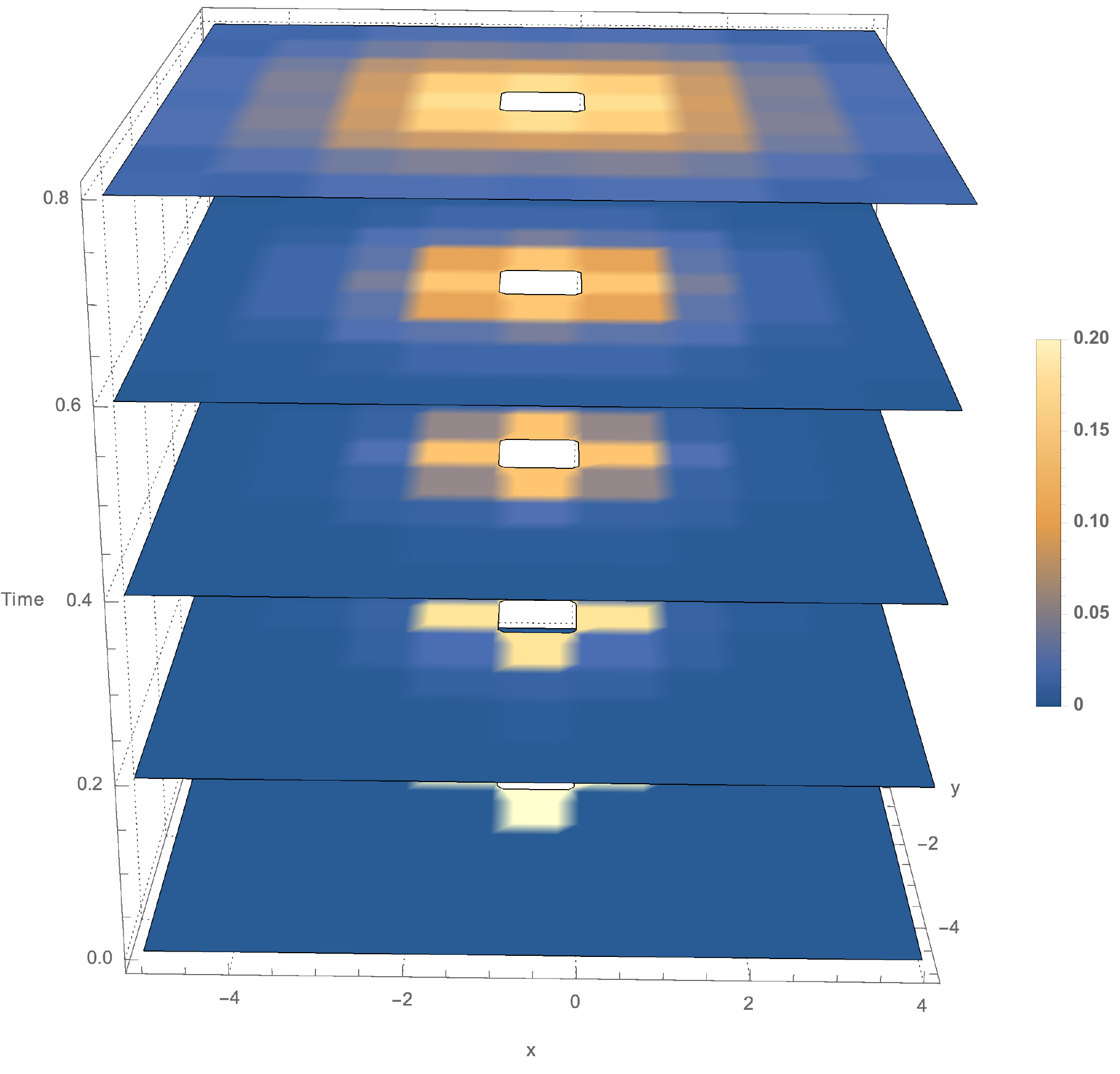} 
 \end{center}
 \caption{
Density plot of $|T_{ij}(t)|$ 
shown as a function of $(x,y)$ at different time slices,
where $x=x_i-x_j$, $y=y_i - y_j$
with 
$x_i = mod (i-1, \sqrt{L} )+1$, $y_i = (i - x_i)/\sqrt{L} + 1$
for the initial condition in \eq{eq:2Dinitial}
with $\epsilon=0.2$,
$R=U=\lambda=1$ and $L=100$.
$|T_{ij}| \sim 0.5$ at $(x,y)=0$ is not shown in the density plot
to increase the contrast for the non-onsite bi-local fields.
 }
 \label{fig:2D_lattice}
 \end{figure}

Next, we consider an example 
where the dimensionality of the emergent local theory 
is greater than one.
For this, we consider an initial state which has a two-dimensional local structure,
\bqa
T_{ij}(0) &=& T_* \delta_{ij}  + \epsilon \delta_{ d_{ij}^{(2)}, 1}, \nn
P_{ij}(0) &= & P_* \delta_{ij},
\label{eq:2Dinitial}
\eqa
where
 $d_{ij}^{(2)} = 
\sqrt{
\left[ i - j  \right]_{\sqrt{L}}^2 
+
\left[ 
\left\| \frac{i-1}{\sqrt{L}} \right\|
- 
\left\| \frac{j-1}{\sqrt{L}} \right\|
\right]_{\sqrt{L}}^2
}$.
In this case, there is no apparent locality for $T_{ij}(t)$
in terms of the one-dimensional coordinate $i$ and $j$.
Instead, it exhibits a local structure associated with the two-dimensional lattice.
Therefore we introduce a coordinate system $(x_i,y_i)$
which is related to site index $i$ via
$x_i = mod (i-1, \sqrt{L} )+1$, $y_i = (i - x_i)/\sqrt{L}+1$.
In  \fig{fig:2D_lattice}, we show how the bi-local fields spread 
over the two-dimensional lattice as a function of time for the initial condition given by \eq{eq:2Dinitial}.
Clearly the spread of entanglement 
follows the locality defined with respect to the
two-dimensional lattice.
This shows that the same Hamiltonian acts 
as a two-dimensional local Hamiltonian 
when applied to states that exhibit two-dimensional local structures,
and as an one-dimensional Hamiltonian 
to states with one-dimensional local structures.
In relatively local theories,
topology, dimension and geometry
are all emergent properties.

 \begin{figure}[ht]
 \begin{center}
 \includegraphics[scale=0.7]{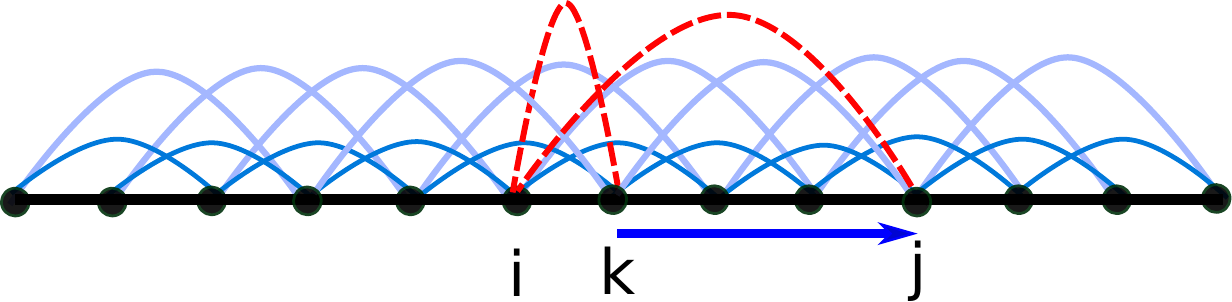} 
 \end{center}
 \caption{
The range of the preformed bi-local fields
(denoted as solid lines)
determines the range of hopping
for a local perturbation added to the system (denoted as dashed lines).
For example, a local disturbance $\delta T_{ik}$
can hop to $\delta T_{ij}$
via the bonds that are formed between 
sites $k$ and $j$. 
 }
 \label{fig:hopping}
 \end{figure}

\section{State dependent propagation of local disturbance}
\label{sec:disturb}

Now we examine 
how a local disturbance propagates in space.
For this, we add an infinitesimally small local perturbation at $i=L/2$
to the initial state in \eq{eq:initial}  
which supports the one-dimensional local structure, 
\bqa
T^Q_{ij}(0) &=& T_* \delta_{ij}  + \epsilon \delta_{ d_{ij}^{(1)}, 1} + Q \delta_{i,L/2} \delta_{j,L/2}, \nn
P^Q_{ij}(0) &= & P_* \delta_{ij},
\label{eq:initial2}
\eqa
where $Q$ denotes the strength of the local perturbation added at site $i=L/2$.
In the limit that  $Q \ll \epsilon$,
the local perturbation propagates
on top of the `geometry' set by the bi-local fields
formed in the absence of the local perturbation. 
We can express the solution in the presence of the local perturbation as
\bqa
T^Q_{ij}(t) & = & T_{ij}(t) +  \delta T_{ij}(t),  \nn
P^Q_{ij}(t) & = & P_{ij}(t) + \delta  P_{ij}(t),
\label{expansion2}
\eqa
where $T_{ij}(t), P_{ij}(t)$ are the solution of the equation of motion with $Q=0$,
and $\delta T_{ij}(t), \delta P_{ij}(t) \sim Q$ represent the deviation generated from the local perturbation.
To the leading order in $Q$, 
$\delta T_{ij}(t), \delta P_{ij}(t)$ satisfy,
\bqa
-i~ \partial_\tau \delta T_{ij} & = &
R \left( - 2 \delta T_{ij} 
+ 8 \delta T_{ik} P_{kl} T_{lj}  
+ 8 T_{ik} \delta P_{kl} T_{lj}  
+ 8 T_{ik} P_{kl} \delta T_{lj}  
\right)    \nn
&& - 4 U 
\left( \delta T_{ik} T_{kj}  
+ T_{ik} \delta T_{kj} 
\right)
+ 2 \lambda  \delta P_{ii}  \delta_{ij}, \nn
i~ \partial_\tau \delta P_{ij} & = &
R \left( - 2 \delta P_{ij} 
+ 8 \delta P_{ik} T_{kl} P_{lj}  
+ 8 P_{ik} \delta T_{kl} P_{lj}  
+ 8 P_{ik} T_{kl} \delta P_{lj}  
\right)   \nn
&& 
-4 U \left( 
   \delta P_{ik} T_{kj}
 + P_{ik} \delta T_{kj}
  +\delta  T_{ik} P_{kj}
 + T_{ik} \delta P_{kj}
 \right).
\label{fulleomQ}
\eqa
Here the bi-local fields that are already formed in the absence of the local perturbation
sets the background on which the perturbation propagates.
The range of the hopping for the perturbation is set by the range of the preformed bi-local fields.
For example,  
$-i~ \partial_\tau \delta T_{ij} \sim
R   \delta T_{ik} P_{kl} T_{lj}  $
in \eq{fulleomQ}
describes the processes in which
$\delta T_{ik}$ defined on link $(i,k)$
jumps to link $(i,j)$
via $ P_{kl} T_{lj}  $
that provides a connection between $k$ and $j$.
If $P_{kl} T_{lj} \sim e^{-d^{(1)}_{kj}/ \xi}$,
$\delta T_{ik}$ jumps by coordinate distance $\xi$ at a time.
This is illustrated in \fig{fig:hopping}. 
Since the profiles of $T_{ij}$ and $P_{ij}$ strongly depend on the initial state, 
so does the coordinate speed at which local
disturbance propagates in space.
As time increases, the range of $T_{ij}$ and $P_{ij}$ increases.
Accordingly, we expect that the coordinate speed for 
the propagation of the local disturbance increases in time.

 \begin{figure}[ht]
 \begin{center}
   \subfigure[]{
 \includegraphics[scale=0.7]{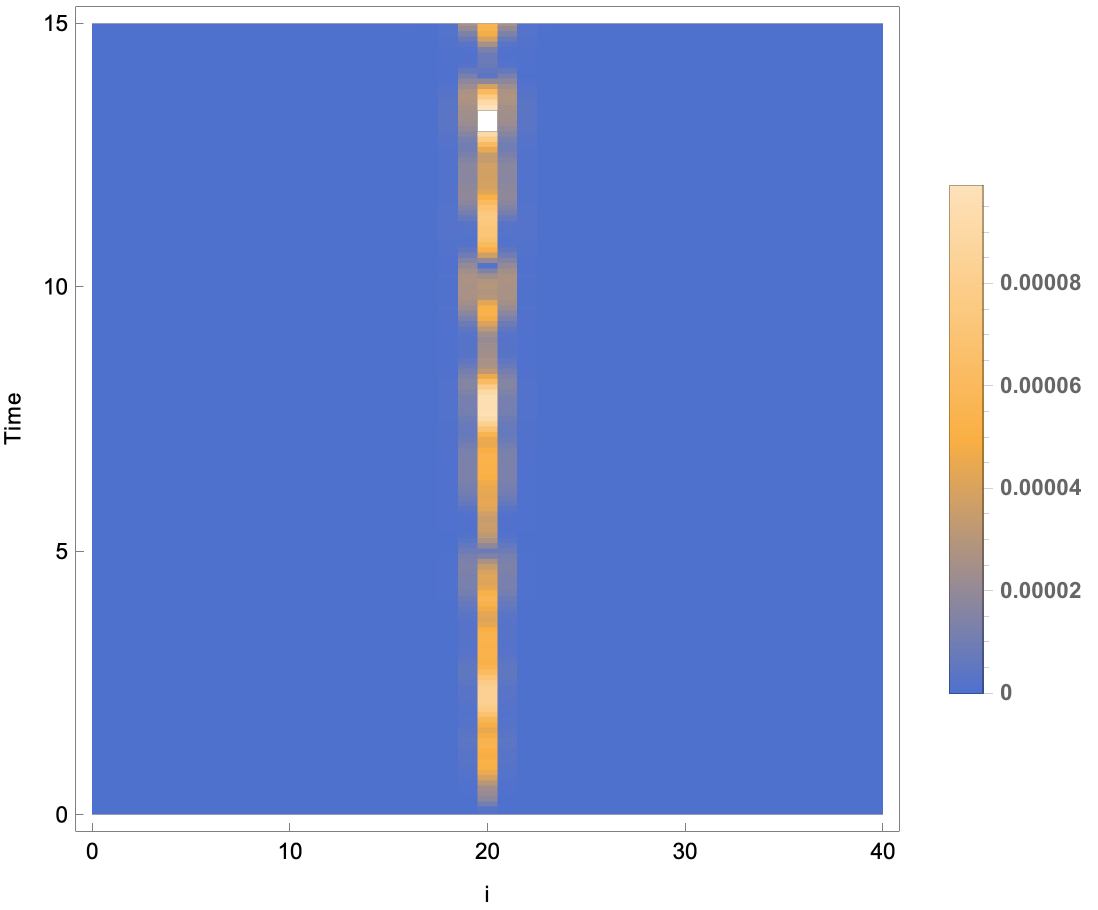} 
 } 
 \hfill
 \subfigure[]{
 \includegraphics[scale=0.7]{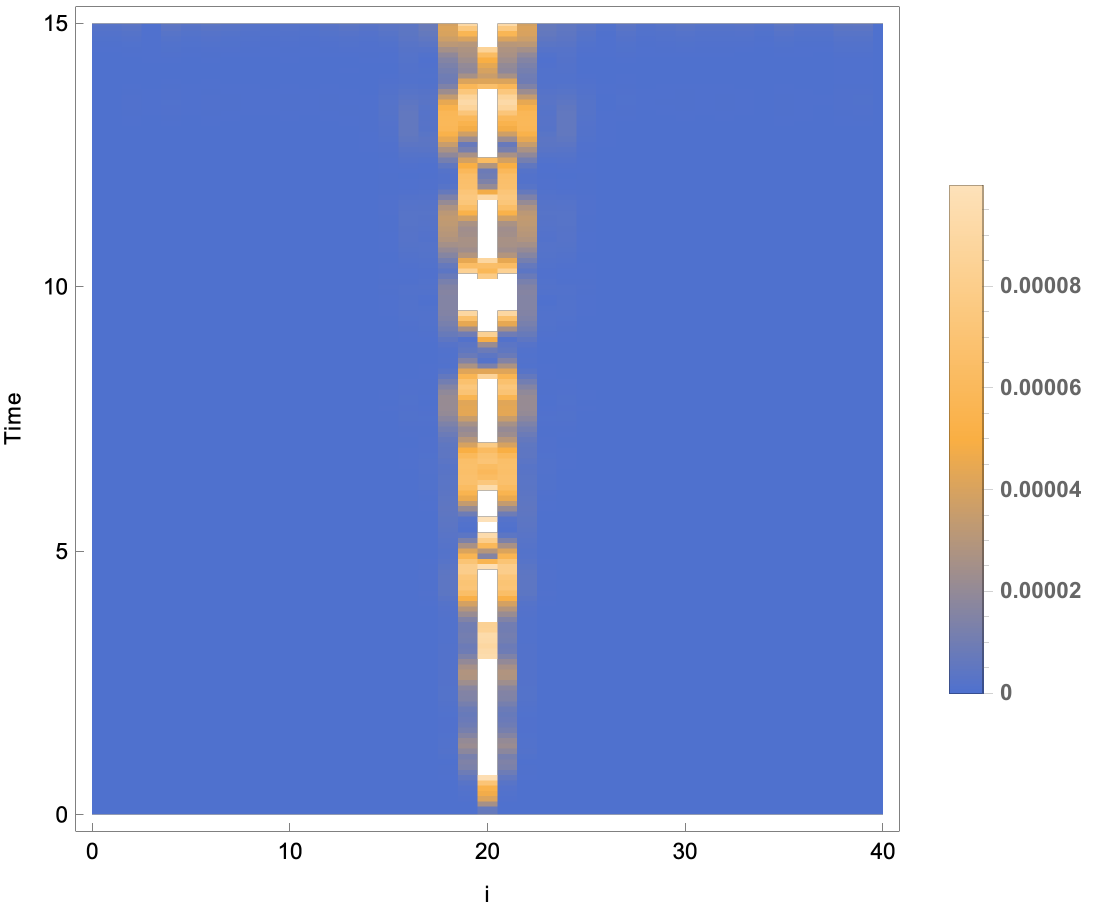} 
 } 
  \subfigure[]{
 \includegraphics[scale=0.7]{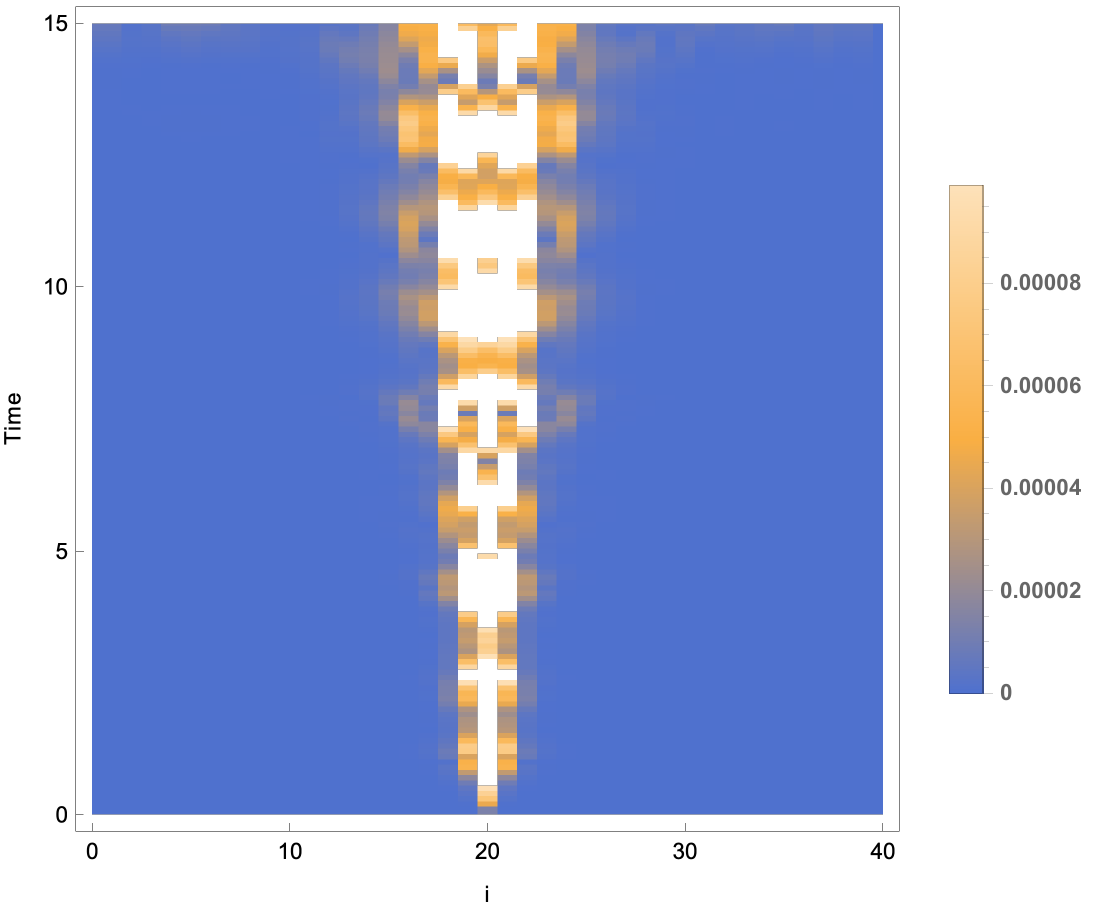} 
 } 
 \hfill
 \subfigure[]{
 \includegraphics[scale=0.7]{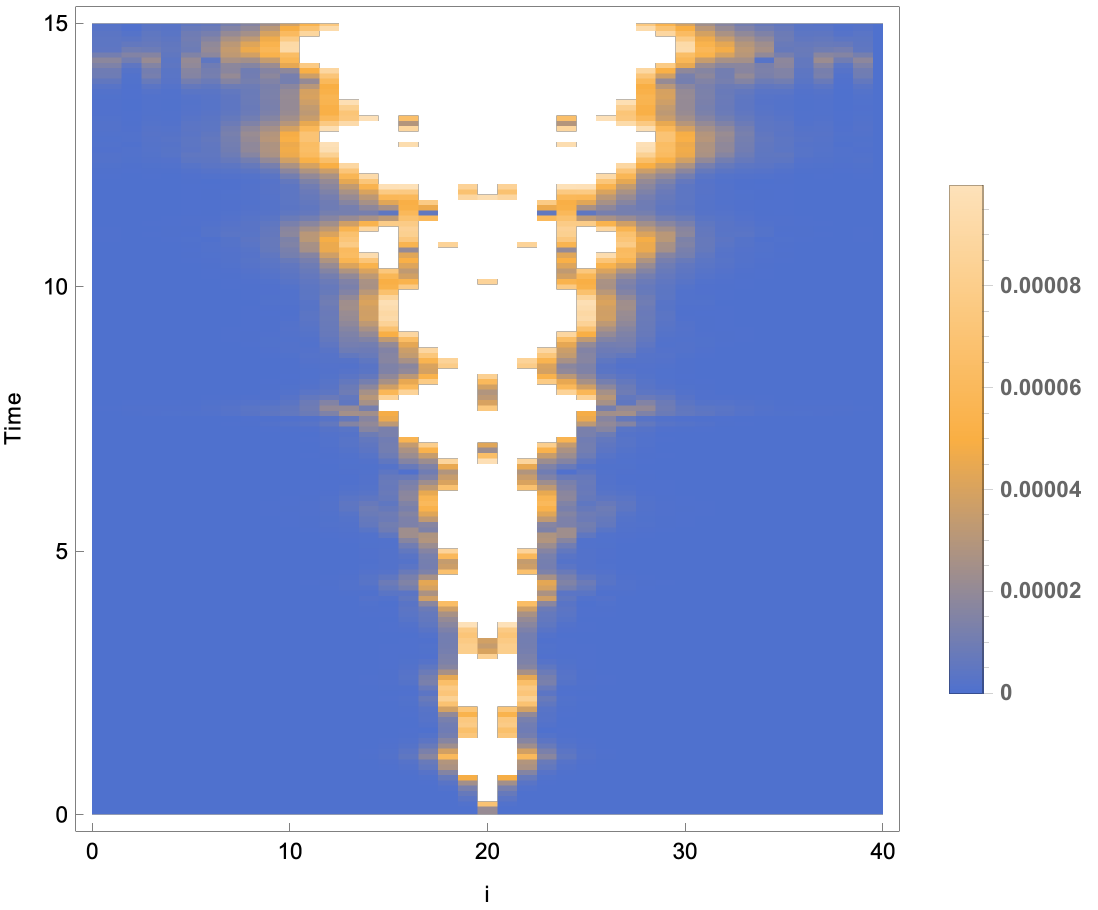} 
 } 
 \end{center}
 \caption{
The relative local energy density $\Delta h_i(t)$ plotted in the space of $i$ and $t$
 for the initial condition in \eq{eq:initial2}
with $R=U=\lambda=1$, $Q=10^{-3}$ and $L=40$
for (a) $\epsilon=0.08$, (b) $\epsilon=0.12$,
 (c) $\epsilon=0.16$ and (b) $\epsilon=0.2$.
 }
 \label{fig:spread_of_perturbation}
 \end{figure}

We check this through the numerical solution obtained for the 
initial condition in \eq{eq:initial2}.
To keep track of how the local perturbation propagate in space,
we introduce
\bqa
h_i 
& = &
R \left( - 2 T_{ii} P_{ii} + 4 (T_{ii} P_{ii} )^2    \right)   
+ U \left( 2 T_{ii} -  4 T_{ii}^2 P_{ii}   \right) 
+ \lambda  P_{ii}^2,
\eqa 
where the index $i$ is not summed over.
This corresponds to the conserved energy density
for the direct product state.
For general states, this is neither conserved nor real.
Nonetheless, this is a useful measure that characterizes 
how far each site is driven away from the ultra-local fixed point.
In \fig{fig:spread_of_perturbation}, we plot
\bqa
\Delta h_i(t) = \left|
\frac{ h_i(t) - h_{L}(t) }{ h_L(t)}
\right|
\label{dh}
\eqa
as a function of $i$ and $t$.
$\Delta h_i(t)$ measures the deviation of the `local energy density' at site $i$ 
relative to the local energy density at a site far away from the site with the local perturbation. 
As is shown in  \fig{fig:spread_of_perturbation},
the speed at which the local disturbance propagates in space strongly depends 
on $\epsilon$.
The coordinate speed of the propagation increases with time as well.
This is due to the fact that the perturbation can hop by larger coordinate distances
using larger entanglement bonds at later time.
In other words, the shrinking universe 
makes coordinate speed of propagation to increase.
The acceleration of the propagating mode in coordinate speed is more manifest
for large $\epsilon$ as is shown in \fig{fig:spread_of_perturbation} (d).
For small $\epsilon$, it is hard to differentiate 
the speed of the propagating mode
from the speed of entanglement spread
because both are small.

It is emphasized that $\epsilon$ is not a parameter
of the Hamiltonian, but a parameter that 
characterizes the amount of entanglement in states.
The state dependent speeds of entanglement spread and wave propagation 
is a hallmark of relatively local Hamiltonians.
The state dependence of {\it coordinate speed} can be understood 
in terms of state dependent geometry with a fixed {\it proper speed}.
A state which is close to a direct product state gives rise to a geometry 
whose proper size is large.
This translates to a small speed of propagation
when the speed is measured in coordinate distance.
Conversely, a state with a larger entanglement 
has a geometry with shorter proper distance, 
and exhibits a larger coordinate speed.
Relatively local Hamiltonians do not satisfy 
a bound on how fast entanglement can spread
in terms of coordinate speed\cite{Lieb1972}.
However, it is likely that there exists a bound on a proper speed,
where the physical distance is measured with respect to state dependent geometry
in the large $N$ limit.

\section{Summary}

In this paper, we construct a simple relatively local Hamiltonian for $N$ scalar fields.
The Hamiltonian is defined on a set of sites 
which has no preferred background.
Although the Hamiltonian is non-local as a quantum operator, 
it acts as a local Hamiltonian in the large $N$ limit
when applied to states whose pattern of entanglement 
exhibits a local structure.
The dimension, topology and geometry of the emergent local theory
are determined by states.
This is shown by examining how states with local structures evolve under 
the time evolution generated by the relatively local Hamiltonian.
The fact that geometry is determined by state manifests itself 
as state-dependent coordinate speeds 
at which entanglement and local perturbation spread.
The state-dependent coordinate speed is naturally understood as 
originating from the state dependent metric.
The metric associated with a saddle-point configuration
can be explicitly determined from the equation of motion
that small fluctuations of the collective variables obey in the continuum limit
if the analytic form of the saddle-point solution is known\cite{Lee2016}.
In the present case, the analytic solution for the strongly time-dependent 
saddle-point solution is not yet available,
and we postpone the derivation of the metric to future work.
The metric determined in this manner is expected to satisfy 
the condition that gapless modes propagate with a fixed proper speed.

Generically, states with local structures undergo `big crunches' 
in which locality is destroyed as sites become globally connected at a later time.
One can also consider a `big bang' 
where locality emerges out of initial states that do not have local structure.
The time reversal symmetry guarantees that there exists
a set of states that exhibit such dynamically emergent locality
at least over a finite window of time. 
Namely, the backward evolution of the big crunches, 
whose evolution is governed by the same Hamiltonian
due to the time reversal symmetry, 
describes an expanding universe where the proper volume of the system
increases as states become less entangled in time.
It would be of great interest to understand 
how generic such states are in the Hilbert space.

\section{A possible roadmap toward quantum gravity}

In relatively local theories,
the metric is nothing but a collective variable 
that determines the pattern of entanglement of underlying quantum matter.
The relatively local theory is 
a background independent theory of geometry
because geometry is dynamically determined 
and there is no fixed background metric.
In this section, we discuss other aspects of gravity
that need to be incorporated into a relatively local theory
so that it becomes a full quantum theory of gravity 
that has a chance to reproduce the general relativity 
in the semi-classical limit.

The key additional ingredient is the diffeomorphism invariance.
The present theory has the discrete spatial diffeomorphism.
However, it does not have the full spacetime diffeomorphism invariance 
because it has a preferred time. 
In the Hamiltonian formulation of Einstein's general relativity, 
spacetime diffeomorphism is represented as the algebra
obeyed by the three momentum and one Hamiltonian densities.
It is likely that not all details of the algebra 
derived from the classical general relativity 
can be carried over to the quantum level.
The precise form of the constraint algebra remains unknown at the quantum level
because of the difficulty in regularizing UV divergences of local operators in the general relativity.
More importantly, the algebra for the four-dimensional spacetime should be viewed 
only as an approximation of a more general algebra 
if the dimension is an emergent property as is the case for relatively local theories.
Nonetheless, the gauge principle itself is likely to survive at the quantum level.
If so, there should exists a closed algebra of constraints
that reduce to diffeomorphisms of manifolds in states with local structures.

In theories where metric emerges as a collective degree freedom,
the underlying quantum matter provides an unambiguous norm in the Hilbert space.
This may give new insights into the problem of time in quantum gravity. 
In principle, physical states should be invariant under all gauge transformations
generated by the constraints.
However, states that are invariant under the constraints are generally 
non-normalizable, and physical states with finite norms spontaneously break 
the symmetry generated by the constraints.
In this framework, a non-trivial time evolution in quantum gravity
can be viewed as a consequence 
of the spontaneously broken gauge symmetry\cite{Lee2018}.

The full theory of collective variables also includes
other dynamical fields besides metric.
For example, the bi-local field considered in this paper
can be decomposed into an infinite tower of local fields with higher spins.
They should contribute to the energy-momentum tensor 
that source the curvature for the spin $2$ mode.
The way metric is coupled with other fields
is dictated by the constraint algebra. 
At the microscopic level, such interactions describe couplings among 
different collective modes of the same underlying matter.
For example, the spin $2$ collective mode that dictates the entanglement 
of the matter can be affected by the spin $0$ collective mode
which describes disturbances localized at sites.

In the tower of modes with general spins, 
there exist gauge fields
if there is a symmetry 
under which the collective variables
carry non-trivial charges.
For example, \eq{t} is invariant under the site-dependent $Z_2$ symmetry,
\bqa
\phi^a_i & \rightarrow & (-1)^{\delta_{ij}} \phi^a_i, \nn
T_{i_1 i_2 .. i_{2n}} & \rightarrow & (-1)^{\sum_{k=1}^{2n} \delta_{i_k j}} T_{i_1 i_2 .. i_{2n}}
\eqa 
for each $j$, and so is \eq{Hhat}. 
As a consequence,   Eqs. (\ref{eq:STP})  and (\ref{eq:onH}) take the form of the $Z_2$ gauge theory
in the temporal gauge, where
the sign of the bi-local field $T_{ij}$ 
becomes the dynamical $Z_2$ gauge field. 
In background independent theories, 
the emergence of gauge fields is general 
beyond the specific group and the one-form gauge field.
For example, in the presence of unbreakable closed loops,
a dynamical two-form U(1) gauge field arises in the bulk\cite{Lee2012}.
This is similar to the way in which global symmetries are promoted to gauge symmetries
in the quantum renormalization group\cite{doi:10.1142/S0217751X13501662}.

Finally, we comment on the relation with other models of emergent locality.
In Refs. \cite{2006hep.th...11197K,PhysRevD.77.104029}, 
models of dynamical graph 
have been considered where a graph with locality 
emerges as a ground state.
The main feature that is shared by the models of quantum graphity and the present model
is that both theories are background independent.
Only a special set of states exhibit locality.
The dimension, topology and geometry are dynamical properties
of those states that have locality.
There are also some important differences.
The first difference lies in the fundamental degree of freedom.
In the graph model, dynamical degrees of freedom are assigned 
to every possible link between vertices. 
Different graphs are formed depending on which links are turned on or off.
In the present theory, on the other hand, 
entanglement between matter fields determine the geometry. 
While connectivity itself is the fundamental degree of freedom in the graph model,
collective variables for underlying local degrees of freedom replaces the role of links in the present model.
In that the fundamental degrees of freedom are ordinary matter fields,
the present approach is more closely related to the way
gravity emerges from field theories in AdS/CFT correspondence\cite{Maldacena:1997re,Witten:1998qj,Gubser:1998bc}.
The fact that metric is a collective variable of quantum matter
in the present model gives rise to a second important distinction from other models.
In the graph models, each vertex can be in principle connected to any number of vertices,
and one needs to introduce an energetic penalty to suppress graphs with too many connections.
In the present model, on the other hand, there is a limited amount of dynamical connections 
a site can form with other sites due to the monogamous nature of entanglement.
While this is not manifest in the large $N$ limit with a small number of sites ($L$),
this constraint is expected to play an important role 
in dynamically suppressing non-local graphs 
in the physical limit with $L \gg N$.
The third difference is the locality of the effective theory
that emerges once a local structure is dynamically selected.
In the present model, the effective theory that describes small fluctuations around 
a saddle-point with local structure is a local theory as is shown in Sec. \ref{sec:disturb}.
This seems less clear in the graph models.
The emergence of a local graph does not necessarily 
guarantee the locality of the effective theory 
that describes small fluctuations of graph 
because the underlying Hamiltonian that selects a local graph is highly non-local.


\section*{Acknowledgments}

The research was supported by
the Natural Sciences and Engineering Research Council of
Canada.
Research at the Perimeter Institute is supported
in part by the Government of Canada
through Industry Canada,
and by the Province of Ontario through the
Ministry of Research and Information.

 \bibliography{references}

\newpage
\begin{appendix}

\section{Method of numerical integration}

In order to solve \eq{fulleom} numerically, we employ 
the symplectic integrator developed for
non-separable Hamiltonian systems\cite{PhysRevE.94.043303}.
\eq{fulleom} is not separable,
and the usual symplectic integrator can not be readily applied.
Therefore, we first double the degrees of freedom and introduce a Hamiltonian
for the enlarged system,
\bqa
{\cal H}_t[T_A, P_A, T_B, P_B] & = & 
{\cal H}[T_A, P_B] + {\cal H}[T_B,P_A] + 
\frac{\omega}{2} \sum_{ij} \left[  (P_{A;ij} - P_{B,ij} )^2 -  (T_{A;ij} - T_{B,ij} )^2   \right]. \nn
\label{Ht}
\eqa 
We solve the equation of motion for the enlarged system 
with the initial condition,  
$T_{A;ij}(0) = T_{B;ij}(0) = \bar T_{ij}$
and 
$P_{A;ij}(0) = P_{B;ij}(0) =\bar P_{ij}$.
The exact solution to the enlarged equation of motion
agrees with the solution to \eq{fulleom}.
The numerical advantage of using \eq{Ht} is that
one can implement the symplectic algorithm
that prevents total energy from drifting as a result
of accumulated numerical error.
We implement an evolution of a state from $t$ to $t+dt$ as
\bqa
\left( \begin{array}{c} T_{A;ij}(t+dt) \\ P_{A;ij}(t+dt) \\ T_{B;ij}(t+dt) \\ P_{B;ij}(t+dt) \end{array} \right)
& = &
\phi^{dt/2}_1
\circ
\phi^{dt/2}_2
\circ
\phi^{dt}_3
\circ
\phi^{dt/2}_2
\circ
\phi^{dt/2}_1
\left( \begin{array}{c} T_{A;ij}(t) \\ P_{A;ij}(t) \\ T_{B;ij}(t) \\ P_{B;ij}(t) \end{array} \right),
\label{eq:syminteg}
\eqa 
where
\bqa
 \phi^{\delta}_1
\left( \begin{array}{c} T_{A;ij} \\ P_{A;ij} \\ T_{B;ij} \\ P_{B;ij} \end{array} \right)
& =& 
\left( 
\begin{array}{c} 
T_{A;ij} \\ 
P_{A;ij} - i \delta \frac{\partial {\cal H}[T_A, P_B] }{\partial T_{A;ij} } \\ 
T_{B;ij} + i \delta \frac{\partial {\cal H}[T_A, P_B] }{\partial P_{B;ij} } \\ 
P_{B;ij} 
\end{array} 
\right),  ~~~~~~~
\phi^{\delta}_2
\left( \begin{array}{c} T_{A;ij} \\ P_{A;ij} \\ T_{B;ij} \\ P_{B;ij} \end{array} \right)
 = 
\left( 
\begin{array}{c} 
T_{A;ij}+ i \delta \frac{\partial {\cal H}[T_B, P_A] }{\partial P_{A;ij} }  \\ 
P_{A;ij} \\ 
T_{B;ij} \\ 
P_{B;ij} - i \delta \frac{\partial {\cal H}[T_B, P_A] }{\partial T_{B;ij} } 
\end{array} 
\right), \nn
\phi^{\delta}_3
\left( \begin{array}{c} T_{A;ij} \\ P_{A;ij} \\ T_{B;ij} \\ P_{B;ij} \end{array} \right)
& = &
\frac{1}{2}
\left( 
\begin{array}{c} 
T_{A;ij}+ T_{B;ij} +  \ccw ( T_{A;ij}- T_{B;ij} ) + i \ssw ( P_{A;ij} - P_{B;ij} ) \\
P_{A;ij}+ P_{B;ij} +  i \ssw ( T_{A;ij}- T_{B;ij} ) +  \ccw ( P_{A;ij} - P_{B;ij} ) \\
T_{A;ij}+ T_{B;ij} -  \ccw ( T_{A;ij}- T_{B;ij} ) - i \ssw ( P_{A;ij} - P_{B;ij} ) \\
P_{A;ij}+ P_{B;ij} - i \ssw ( T_{A;ij}- T_{B;ij} ) - \ccw ( P_{A;ij} - P_{B;ij} ) 
\end{array} 
\right). \nn
\eqa
The symplectic form 
$dT_{A;ij} \wedge dP_{A;ij} + dT_{B;ij} \wedge dP_{B;ij}$
is preserved under \eq{eq:syminteg}.
In this paper, we use 
$R=U=\lambda=1$, 
$\omega=10$ and $dt = 10^{-3}$.

\end{appendix}

\end{document}